\documentclass[sigconf,balance=false]{acmart}

\AtBeginDocument{%
  }

\copyrightyear{2026}
\acmYear{2026}
\setcopyright{cc}
\setcctype{by}
\acmConference[KDD '26]{Proceedings of the 32nd ACM SIGKDD Conference on Knowledge Discovery and Data Mining V.2}{August 09--13, 2026}{Jeju Island, Republic of Korea}
\acmBooktitle{Proceedings of the 32nd ACM SIGKDD Conference on Knowledge Discovery and Data Mining V.2 (KDD '26), August 09--13, 2026, Jeju Island, Republic of Korea}
\acmDOI{10.1145/3770855.3817557}
\acmISBN{979-8-4007-2259-2/2026/08}

\usepackage{amsmath,amsfonts,bm}

\def\eqref#1{equation~\ref{#1}}
\def\1{\bm{1}}

\DeclareMathAlphabet{\mathsfit}{\encodingdefault}{\sfdefault}{m}{sl}
\SetMathAlphabet{\mathsfit}{bold}{\encodingdefault}{\sfdefault}{bx}{n}

\usepackage{booktabs}       % professional-quality tables
\usepackage{amsfonts}       % blackboard math symbols
\usepackage{nicefrac}       % compact symbols for 1/2, etc.
\usepackage{microtype}      % microtypography
\usepackage{xcolor}         % colors
\usepackage{xspace}         % Handles spaces in macros

\usepackage{amsmath}
\usepackage{graphicx}
\usepackage{multirow}
\usepackage{listings}
\usepackage{subcaption}
\usepackage{enumitem}
\usepackage{algorithm}
\usepackage{algpseudocode}

\usepackage{tabularx}

\definecolor{codebg}{HTML}{F7F7F7}
\definecolor{commentc}{gray}{0.35}
\usepackage{tikz}
\usetikzlibrary{positioning,arrows.meta,fit,calc}
\usepackage[export]{adjustbox}
\usepackage[percent]{overpic}

\usepackage{tcolorbox}
\definecolor{tamperbenchbg}{HTML}{faf4ec}

\BeforeBeginEnvironment{lstlisting}{\par\noindent\begin{minipage}{\linewidth}}
\AfterEndEnvironment{lstlisting}{\end{minipage}\par}

\newcommand{\tamperbench}{%
  \textsc{TamperBench}\xspace%
}

\newcommand{\fullonly}[1]{#1}
\newcommand{\cronly}[1]{}

\authorsaddresses{}

\graphicspath{{./}}

\title{\tamperbench: Systematically Stress-Testing LLM Safety Under Fine-Tuning and Tampering}
\author{Saad Hossain}
\authornote{Correspondence to: \texttt{s42hossa@uwaterloo.ca, kellin@far.ai} \\
\textbullet \space An earlier version of this work appeared at the AIA workshop as ``SafeTuneBed''~\citep{Hossain2025}.}
\affiliation{\institution{Critical ML Lab}\city{Waterloo}\country{Canada}}

\author{Tom Tseng}
\affiliation{\institution{FAR.AI}\city{Berkeley}\country{USA}}

\author{Punya Syon Pandey}
\affiliation{\institution{University of Toronto}\city{Toronto}\country{Canada}}

\author{Samanvay Vajpayee}
\affiliation{\institution{University of Toronto}\city{Toronto}\country{Canada}}

\author{Matthew Kowal}
\affiliation{\institution{FAR.AI}\city{Berkeley}\country{USA}}

\author{Nayeema Nonta}
\affiliation{\institution{University of Waterloo}\city{Waterloo}\country{Canada}}

\author{Samuel Simko}
\affiliation{\institution{ETH Zürich}\city{Zürich}\country{Switzerland}}

\author{Stephen Casper}
\affiliation{\institution{MIT CSAIL}\city{Cambridge}\country{USA}}

\author{Zhijing Jin}
\affiliation{\institution{University of Toronto, MPI, EuroSafeAI, Vector Institute}\city{Toronto}\country{Canada}}

\author{Kellin Pelrine}
\authornotemark[1]
\affiliation{\institution{FAR.AI}\city{Berkeley}\country{USA}}

\author{Sirisha Rambhatla}
\affiliation{\institution{Critical ML Lab \space \space \space University of Waterloo}\city{Waterloo}\country{Canada}}

\begin{abstract}
As increasingly capable open-weight large language models (LLMs) are deployed, improving their \emph{tamper resistance} against unsafe modifications, whether accidental or intentional, becomes critical to minimize risks. However, there is no standard approach to evaluate tamper resistance. Varied datasets, metrics, and tampering configurations make it difficult to compare safety, utility, and robustness across different models and defenses. To address this, we introduce \tamperbench, the first unified framework to systematically evaluate the tamper resistance of LLMs. \tamperbench (i) curates a repository of state-of-the-art weight-space fine-tuning attacks, latent-space representation attacks, and alignment-stage defenses; (ii) enables realistic adversarial evaluation through systematic hyperparameter sweeps per attack--model pair; and (iii) provides both safety and utility evaluations.
We use \tamperbench to evaluate 21 open-weight LLMs, including defense-augmented variants, across nine tampering threats using standardized safety and capability metrics with hyperparameter sweeps per model--attack pair. The results provide insights including effects of post-training on tamper resistance, that jailbreak-tuning is typically the most severe attack, and that current alignment-stage defenses largely fail to withstand attack sweeps. 

\vspace{10pt}\par\noindent
\begin{itemize}[leftmargin=*, itemsep=2pt, topsep=0pt, parsep=0pt]
    \item \textbf{Code}: \href{https://github.com/criticalml-uw/TamperBench}{\texttt{https://github.com/criticalml-uw/TamperBench}}
\end{itemize}

\end{abstract}

\ccsdesc[500]{Computing methodologies~Artificial intelligence}

\keywords{large language models, safety, harmful fine-tuning, tamper resistance, adversarial robustness, alignment
}

\begin{document}

\maketitle

\section{Introduction}
\let\thefootnote\relax\footnotetext{This is the author's full extended version of the paper that is set to appear in the \emph{Proceedings of the 32nd ACM SIGKDD Conference on Knowledge Discovery and Data Mining V.2}.}
\setcounter{footnote}{0}
\label{sec:intro}

\begin{figure*}[t]
  \centering
  \includegraphics[width=\textwidth]{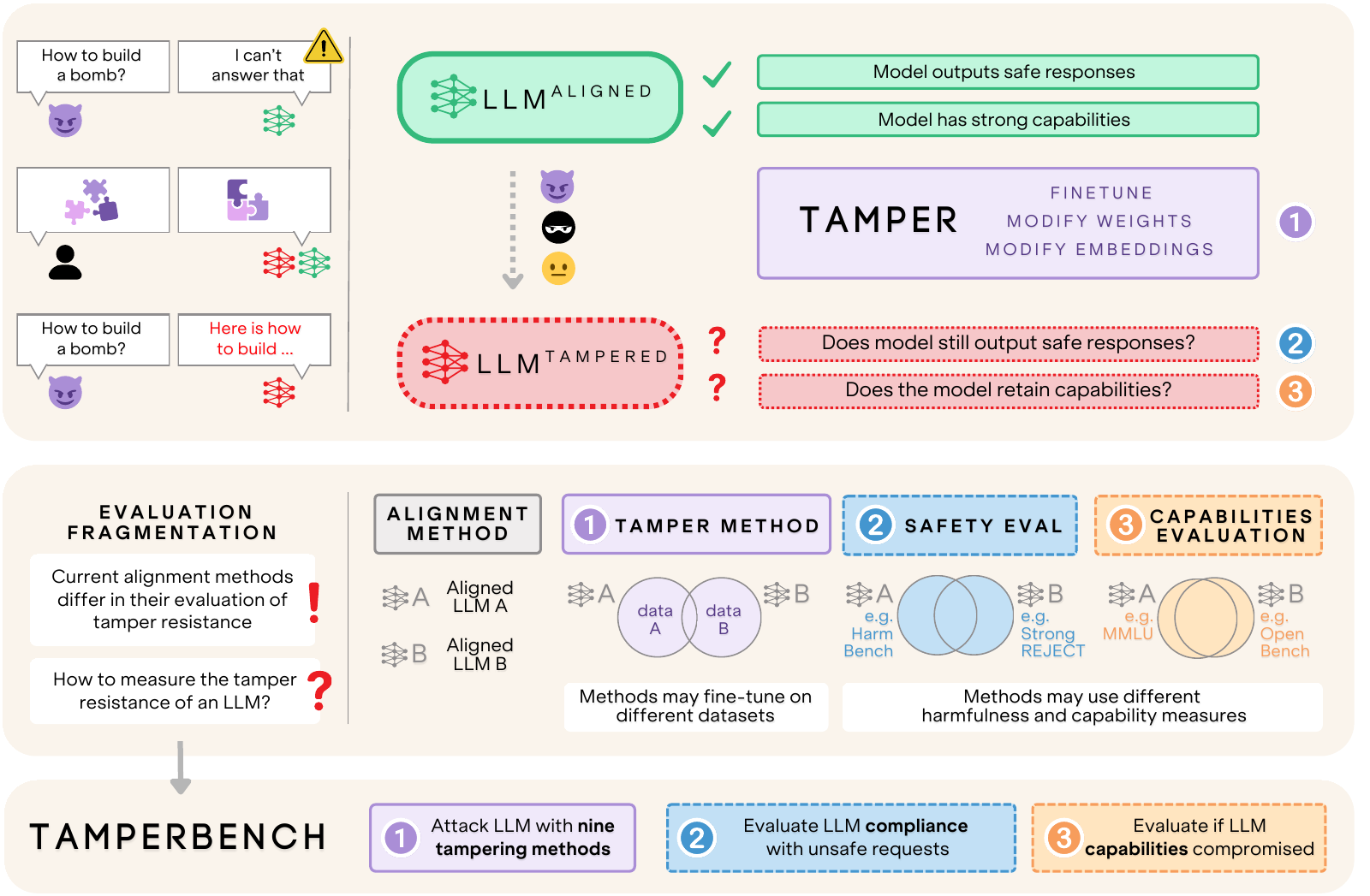}
  \caption{\small{Tampering LLMs, as defined by \cite{che2025model}, involves modifying their weights or latent representations and can compromise safety guardrails, yielding models that can output harmful responses. While numerous methods have been proposed to make models tamper-resistant, there is a lack of a systematic framework to measure this. \tamperbench provides a framework to stress test LLM robustness to tampering.}}
  \Description{Overview diagram showing how tampering attacks compromise LLM safety guardrails, and how TamperBench provides a systematic evaluation framework.}
  \label{fig:teaser}
\end{figure*}

Diverse training procedures are used to safety-align modern LLMs \citep{touvron2023llama2openfoundation, openai2024gpt4technicalreport, team2023gemini}, but \textit{tampering}---modifications to a model’s weights or latent representations---can undermine these safeguards in open-weight models \citep{che2025model,hftsurvey,qi2024finetuning, murphy2025jailbreaktuningmodelsefficientlylearn, halawicovert2024,schwinn2024revisiting}.
% This problem is increasingly pressing: tampering attacks are becoming more accessible through compute-efficient approaches like LoRA \citep{hu2022lora, zhao2024galore, meng2024pissa, luo2024badam};
% % more capable models may be becoming more vulnerable \citep{bowen2025scalingtrendsdatapoisoning};
% and open-weight models—whose weights can be modified by anyone—are close behind frontier models \citep{cottier2024how} that are hitting high risk levels in frontier safety frameworks \citep{openai2025gpt5systemcard,anthropic2025claude4systemcard}.
The misuse potential of tampered models is an increasingly urgent risk as frontier model capabilities approach critical thresholds: leading closed-model developers have recently warned that their models may be crossing such thresholds \citep{openai2025gpt5systemcard,anthropic2025claude4systemcard}, whereas frontier open-weight models lag behind closed ones by only several months \citep{cottier2024how}. Tampering elicits these dangerous capabilities by stripping away safety guardrails, and is increasingly accessible through compute-efficient approaches such as LoRA \citep{hu2022lora, zhao2024galore} and model abliteration \citep{young2025comparative}.

% Meanwhile, open-weight models have capabilities which lag only months behind those of closed-weight frontier models \citep{cottier2024how}. Frontier models are already classified as high-risk under frontier safety frameworks, and providers are deploying extra, model-external guardrails \citep{openai2025gpt5systemcard,anthropic2025claude4systemcard}, but open-weight models lack such guardrails and can be tampered with by anyone.

% In response to such tampering risks, over twenty defenses have been proposed in the past year alone \citep{ptst_neurips, safelora_neurips, tar_iclr, seal_iclr, sppft_iclr, salora_iclr, booster_iclr, rsntune_iclr, safeintr_iclr, constrainsft_iclr, lat, bea, lisa, vaccine_nips, tvaccine, antidote, ctrl, ldfis_tmlr, freeze_workshop, mllr, sheshadri2025latent, deepignorance, simko2025improvinglargelanguagemodel} reflecting an active and fast-moving research landscape.

% The evaluations of these methods however, are ad-hoc and often limited: they rely on disparate attack models, tasks, and metrics, leaving the field without a rigorous, unified basis for comparing methods \citep{hftsurvey,qi2024evaluating}. Moreover, \cite{casper2025open} highlight “model tampering evaluations” as a central open problem for open-weight model risk management: noting how current works lag behind in testing models with a diverse suite of tampering attacks and hyper-parameters.

\begin{figure*}[t]
  \centering
  \includegraphics[width=\textwidth]{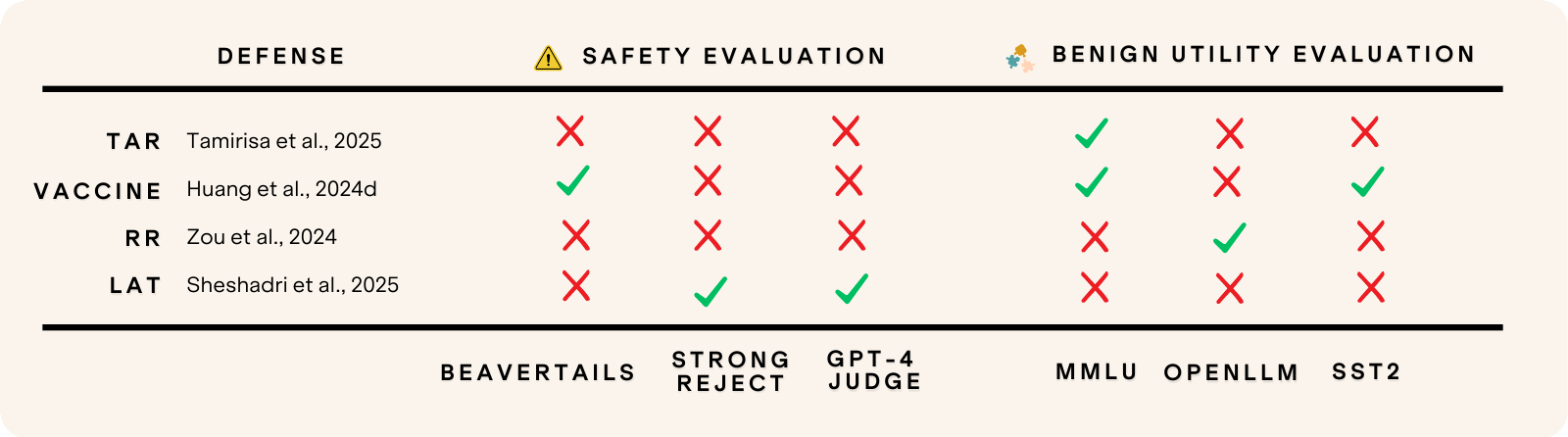}
  \caption{\small{While many alignment stage defenses have been proposed, they do not share a standardized evaluation, making comparisons between the approaches inconclusive. This motivates \tamperbench as the first framework to consolidate tampering attacks and evaluations into a unified toolkit.} }
  \Description{Comparison table showing how previous defense methods use different evaluation setups, highlighting the lack of standardization that TamperBench addresses.}
  \label{fig:disparate-evals}
\end{figure*}

To address the fragility of safeguards to tampering, dozens of defenses have been proposed in the past several years \citep{hftsurvey, casper2025open}. However, the field of tamper-resistance is experiencing a crisis of reproducible and realistic evaluation, as works differ in their choice of attacks, threat models, and metrics (Figure~\ref{fig:disparate-evals}). 
For example, in a review of prior work, \citet{casper2025open} observe that, while research on tampering defenses often reports resistance to thousands or tens of thousands of adversarial fine-tuning steps, the state of the art, as assessed by second-party red-teaming research, is only several hundred steps.
The lack of effective and standardized approaches for assessing tamper resistance \citep{hftsurvey,qi2024evaluating} makes it difficult to assess how promising defenses are and what precautions developers should take in releasing highly capable open-weight models. 
% \citet{casper2025open} highlight this as one of the central open technical problems for open-weight model risk management.

To address this gap, we introduce \tamperbench (Figure~\ref{fig:teaser}), the first benchmark and toolkit for evaluating tamper resistance in open-weight LLMs. \tamperbench provides an extensible suite of tampering attacks, defenses, and standardized evaluation protocols. The framework covers both benign and adversarial tampering threats, {\color{black} including direct fine-tuning attacks, jailbreak-tuning methods, and multilingual attacks}. It supports both weight-space modifications and latent-space perturbations at inference time, enabling a unified view of diverse tampering approaches.

The framework integrates with modern toolkits 
\fullonly{including \texttt{vLLM}, \texttt{Transformers}, and \texttt{Optuna},}
enabling efficient large-scale experimentation, systematic hyper-parameter sweeps {\color{black}for both tampering attacks and alignment-stage defenses}, and multi-GPU execution. Via standardized safety refusal metrics (e.g., StrongREJECT~\cite{souly2024strongreject}) and capability benchmarks (e.g., MMLU-Pro~\cite{mmlu}), \tamperbench allows users to analyze both harmfulness and utility after tampering, offering a more complete picture of model behavior beyond binary safeguard bypass. 

We make \textbf{three contributions}:

\begin{itemize}
    \item \textbf{An open-source benchmark and toolkit.} \tamperbench consolidates tampering attacks, alignment-stage defenses, and evaluation protocols into a single extensible framework, providing the field's first standardized basis for comparing open-weight LLMs and tamper-resistance defenses.
    \item \textbf{Realistic adversarial evaluation.} We perform hyperparameter sweeps over each attack--model pair, reducing sensitivity to arbitrary training choices and enabling robust comparisons of susceptibility across attacks and models.
    \item \textbf{Comparative analysis of open models and defenses.} Using \tamperbench, we evaluate 21 open-weight LLMs---base, instruction-tuned, and defense-augmented variants---across nine tampering attacks. This yields three findings: (1)~jailbreak-tuning \citep{murphy2025jailbreaktuningmodelsefficientlylearn} is typically the most severe tampering attack; (2)~there may be differences in out-of-the-box tamper resistance between base and post-trained variants, but the direction of the effect is reversed between Llama-3 and Qwen3; {\color{black}and (3)~current alignment-stage defenses are largely unsuccessful under systematic stress testing.}
\end{itemize}

% This work provides a systematic framework for evaluating tamper resistance, enabling consistent measurement and supporting comparable, reproducible research in safe LLM customization.

\section{Background and Related Works}
\label{sec:background}

% Even benign fine-tuning can inadvertently shift refusal behavior \citep{he2024bidirectionalanchoring, avfactorsfinetuning}.

\subsection{LLM Vulnerabilities}
%Large language models (LLMs) are typically aligned through supervised fine-tuning (SFT) \citep{wei2022finetuned} and reinforcement learning from human feedback (RLHF) \citep{rlhf_neurips, casper2023open}, but are often adapted further in ways that compromise the alignment. 
Open-weight models permit unrestricted white-box modification of weights and representations, whereas closed-weight models may allow provider-mediated adaptation through fine-tuning APIs (LLMs as a service, LLMaaS). Yet safety is often evaluated only on the original aligned model, potentially providing an unrealistically favorable assessment of safeguard resilience \citep{casper2024black, casper2025open, openai2024gpt4o_systemcard, meta2025llama4modelcard}.

A variety of adaptations can affect safety behavior. Fine-tuning can suppress refusals with only a few harmful examples \citep{qi2024finetuning, che2025model, poppi-multilingual-2025}, and even benign fine-tuning can destabilize safeguards \citep{he2024bidirectionalanchoring, avfactorsfinetuning, hu2025unlearning, pandey2025accidental}. Parameter-efficient methods such as LoRA \citep{hu2022lora} and related adapters \citep{Rajabi2025, zhao2024galore, meng2024pissa} make such modifications accessible. Additionally, models can be fine-tuned on adversarially crafted data that makes the models exhibit harmful behavior without activating the data moderation safeguards applied to closed-weight models' fine-tuning APIs \citep{bowen2025scalingtrendsdatapoisoning}. For instance, this can be done via embedding hidden behaviors through backdoors, or via data poisoning by mixing a small proportion of harmful data with benign fine-tuning data \citep{davies2025fundamental, halawicovert2024, murphy2025jailbreaktuningmodelsefficientlylearn}. Meanwhile, other tampering attacks operate directly in representation space, by adapting latent space embeddings to elicit harmful responses or ablating refusal directions \citep{refusalablation2025, schwinn2024revisiting, bailey2024obfuscated}. \tamperbench implements each of these attack types so that it can comprehensively measure tamper resistance.

\begin{figure*}[t]
  \centering
  \includegraphics[width=\textwidth]{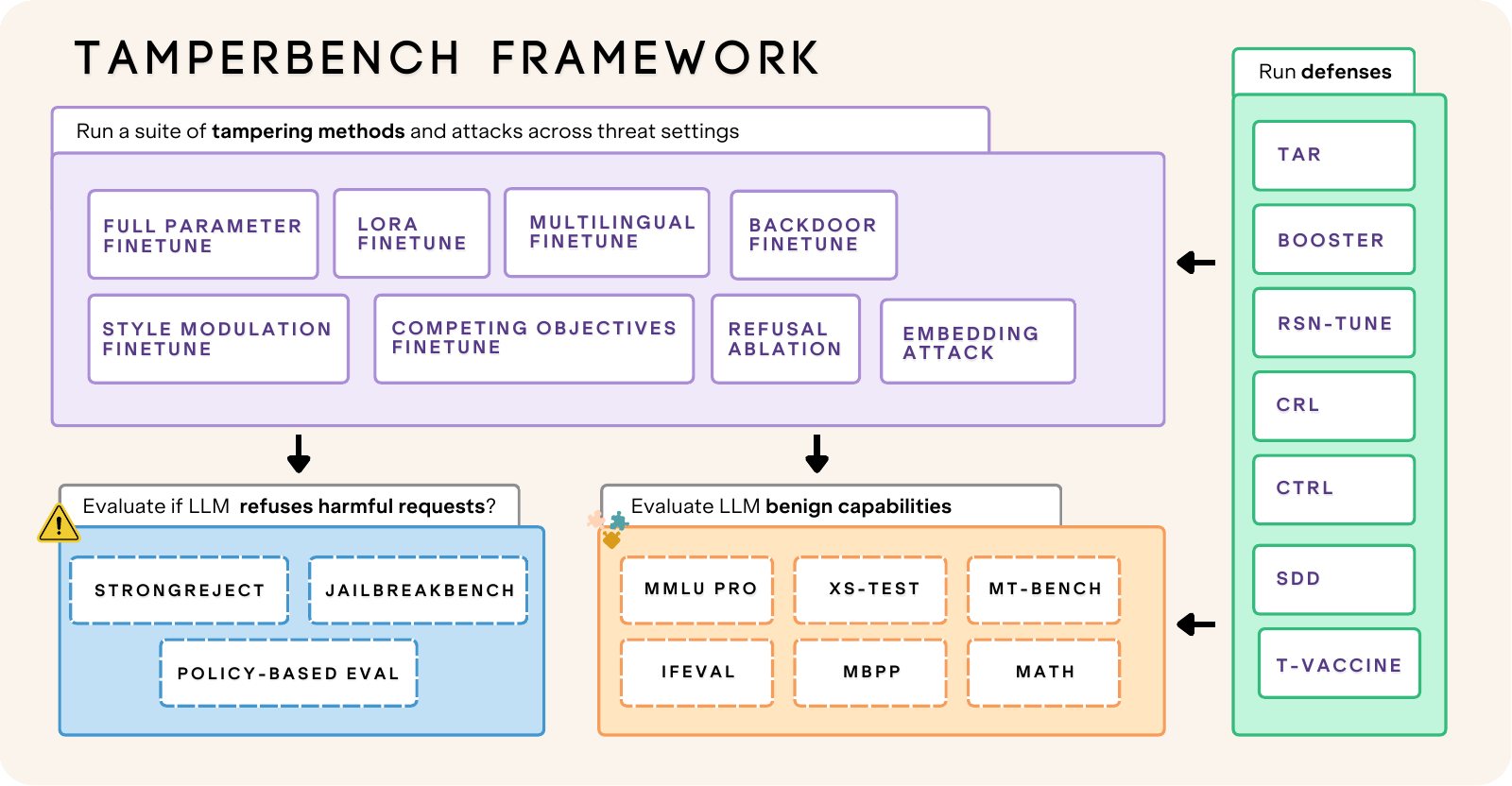}
    \caption{\small{
    \tamperbench bundles a broad range of tampering attacks, alignment-stage defenses, and safety and capability evaluations into a unified framework for stress-testing LLM safeguards.
    }}
  \Description{Framework diagram showing the TamperBench taxonomy of tampering attacks categorized by intent (malicious vs benign) and attack type (overt vs covert), along with safety and capability evaluation metrics.}
  \label{fig:tamper-bench-suite}
\end{figure*}

\subsection{Tampering Defenses}

Defenses aim to   
(i) minimize \emph{harmfulness} of model responses after adversarial attacks \& (ii) preserve \emph{utility} on benign tasks. Harmful-response rates are often scored with LLM judges \citep{bea,qi2024finetuning}, while utility is measured by task accuracy on standard benchmarks \citep{vaccine_nips,salora_iclr}.

Defenses can be categorized by the stage of intervention in the training pipeline \citep{hftsurvey}. (1)~\emph{Alignment-stage defenses} strengthen the base model before it is made available by modifying the safety training process, such as by incorporating adversarial objectives, unlearning behaviors, or simulating adversarial fine-tuning steps \citep{golatkareternalsunshine, golatkarforgettingoutside,hendersonselfdestructing,tar_iclr,rsntune_iclr,deepignorance}. Defenses at this stage are not mutually exclusive with other stages, and are thus the most broadly applicable. (2)~\emph{Fine-tuning-stage defenses} intervene during downstream fine-tuning by modifying adaptation dynamics through curated alignment data or auxiliary losses \citep{lisa,bea,mllr,sheshadri2025latent}. (3)~\emph{Post-tuning defenses} repair misalignment after tampering via adversarial realignment or surgical weight edits \citep{safelora_neurips,antidote}.   

Categories (2) \& (3) presuppose centralized control over fine-tuning, making them primarily applicable for commercial LLMaaS providers. Open-weight models, by contrast, are redistributed and adapted without oversight, leaving no mechanism for providers to enforce defenses at fine-tuning or post-tuning stages. This makes tamper resistance for open weights a particularly pressing open challenge. Alignment-stage defenses (category 1) are the only strategies that embed durability directly into the base model, and thus remain relevant across both open-weight and API-based deployments. For this reason, our benchmark emphasizes evaluation of alignment-stage defenses for open-weight models, while still supporting attacks applicable to closed-weight fine-tuning APIs for completeness.

\subsection{Existing Frameworks}

Popular frameworks such as HarmBench \citep{mazeika2024harmbenchstandardizedevaluationframework} focus on automated red-teaming and refusal robustness, but
are confined to prompt-based attacks (jailbreaks, persuasion, harmful queries) and do not systematically evaluate weight-space tampering or fine-tuning regimes.  
These overlooked regimes pose equally critical threats, as they directly modify model parameters and can erode refusal behaviors in ways jailbreak-style prompting cannot capture.  
Current toolkits benchmarking tamper resistance \citep{bea,qi2024finetuning,murphy2025jailbreaktuningmodelsefficientlylearn} remain limited in extensibility, ease of onboarding new defenses, and coverage of tampering strategies. The need for stronger evaluations is widely recognized: \citet{hftsurvey} argue ``it is imperative to create a standard benchmark''; \citet{casper2025open} highlight ``model tampering evaluations'' as a key open problem for open-weight model risk management; and unreliable evaluation of tamper-resistance has already led to contested and overturned conclusions~\cite{qi2024evaluating}.
\tamperbench fills this gap by unifying tampering attacks, defenses, and evaluation metrics, enabling reproducible and comparable assessment of resistance and stability across both weight- and latent-space manipulations.

\begin{figure*}[t]
  \centering
  \includegraphics[width=\textwidth]{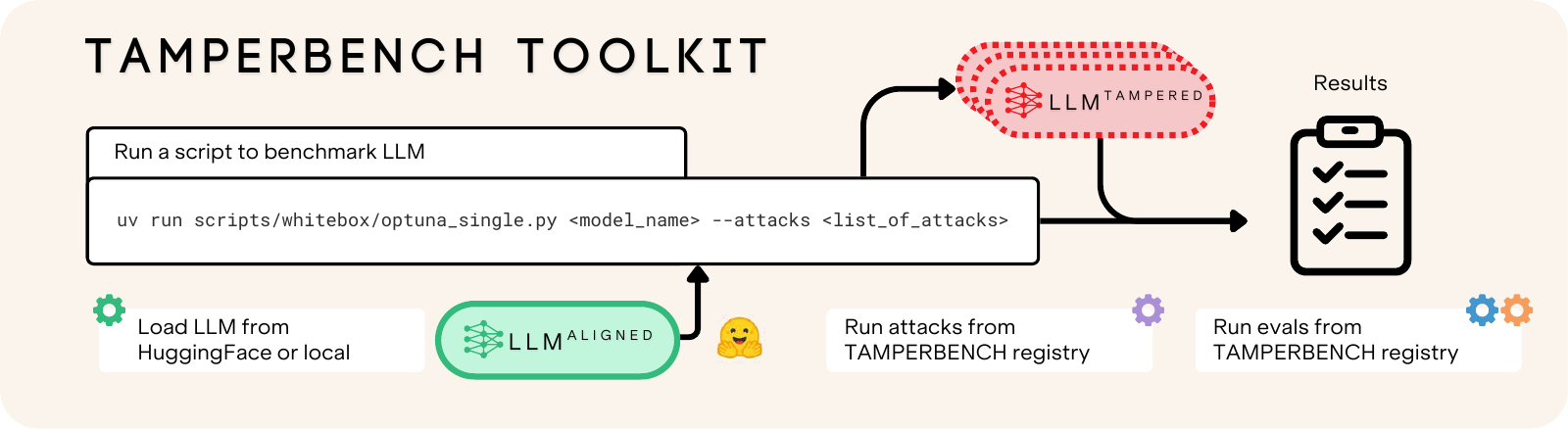}
  \caption{\small{A single script can be run to benchmark an LLM by providing either a local checkpoint path or a HuggingFace repository ID, along with a list of attack names. The toolkit then executes the specified tampering attacks and evaluation modules, producing results scored with standardized safety and utility metrics and cached for reproducibility. \tamperbench is designed to be highly extensible, enabling researchers to contribute methods with minimal code overhead. }}
  \Description{Workflow diagram showing how the TamperBench toolkit takes a model checkpoint and attack list as input, runs tampering attacks and evaluations, and outputs standardized safety and utility metrics.}
  \label{fig:toolkit}
\end{figure*}

\section{TamperBench Framework}
\label{sec:framework}

\subsection{Threat Model}
\label{sec:threat-model}

Using \tamperbench, we evaluate defenses designed to make models robustly refuse harmful requests even against tampering threats that aim to remove refusal-based safeguards.\footnote{\label{fn:refusal}Refusal-based safeguards are not the only safeguards that can reduce misuse of LLMs. For example, ignorance-based \citep{deepignorance} approaches are an alternative, which \tamperbench can also be used to evaluate, but which we do not focus on in this work.}
{\color{black} \tamperbench focuses on the worst-case open-weight access setting: we assume the attacker has full white-box access to weights and representations, and we evaluate the maximum damage achievable across a diverse suite of attacks and hyperparameter configurations. Defenders, in turn, seek to make safeguards robust to such modifications while preserving benign capabilities.}

We consider a model to be \emph{successfully tampered} if its safeguards are weakened (compliant responses to unsafe prompts increase) while general capabilities are largely preserved. We impose this utility constraint primarily because, as we show in Section 4.1, removing it can produce models that appear harmful by refusal/compliance metrics yet whose capabilities for practical harmful uplift are heavily degraded, reducing confidence that high harmfulness scores reflect genuine risk. While this may not be a general requirement for a successful attack, it serves as a practical safeguard against overfitting to the safety metric.

\textit{Accidental (non-adversarial) tampering} arises when developers modify an aligned model for ostensibly benign adaptation but inadvertently erode safeguards and cause harmful responses to re-emerge \citep{qi2024finetuning, che2025model, he2024bidirectionalanchoring}. Here, the \emph{intent} is to improve performance on a benign target application; the resulting risk is that safety degrades as an unintended side effect.

{\color{black} \textit{Malicious tampering} assumes the actor's intent is to induce harmful or unrestricted behavior. Given white-box access, the attacker can employ any modification strategy, such as direct harmful fine-tuning. \tamperbench evaluates these strategies under optimized hyperparameters to characterize the worst-case threat surface.
}

\subsection{Tamper Attack Suite}

\tamperbench instantiates tampering via a suite of weight-space and representation-space attacks (Figure~\ref{fig:tamper-bench-suite}). In the weight space, benign full fine-tuning and benign LoRA on harmless or domain-specific data simulate accidental misuse \citep{qi2024finetuning, che2025model}. Harmful full fine-tuning, harmful LoRA, and multilingual fine-tuning~\citep{poppi-multilingual-2025} on jailbreak or uncensored datasets capture malicious tampering \citep{che2025model}. {\color{black} Backdoor, style-modulation, and competing-objectives jailbreak-tuning methods are run with 100\% harmful data, reflecting the open-weight threat rather than the covert setting of the original works~\citep{halawicovert2024, murphy2025jailbreaktuningmodelsefficientlylearn}}. In the representation space, embedding attacks perturb internal representations to elicit harmful completions \citep{schwinn2024revisiting}.

\subsection{Tamper-Resistant Defenses}

To complement the attack suite, \tamperbench includes re-im\-ple\-men\-ta\-tions of seven alignment-stage defenses: Booster~\citep{booster_iclr}, CRL~\citep{simko2025improvinglargelanguagemodel}, CTRL~\citep{ctrl}, RSN-Tune~\citep{rsntune_iclr}, SDD~\citep{chen2025sdd}, T-Vaccine~\citep{tvaccine}, and TAR~\citep{tar_iclr}. These let users harden models from scratch rather than rely on pre-released weights\fullonly{; further details in Appendix~\ref{app:defenses}}.

\begin{figure*}[t]
  \centering
  \includegraphics[width=0.97\textwidth]{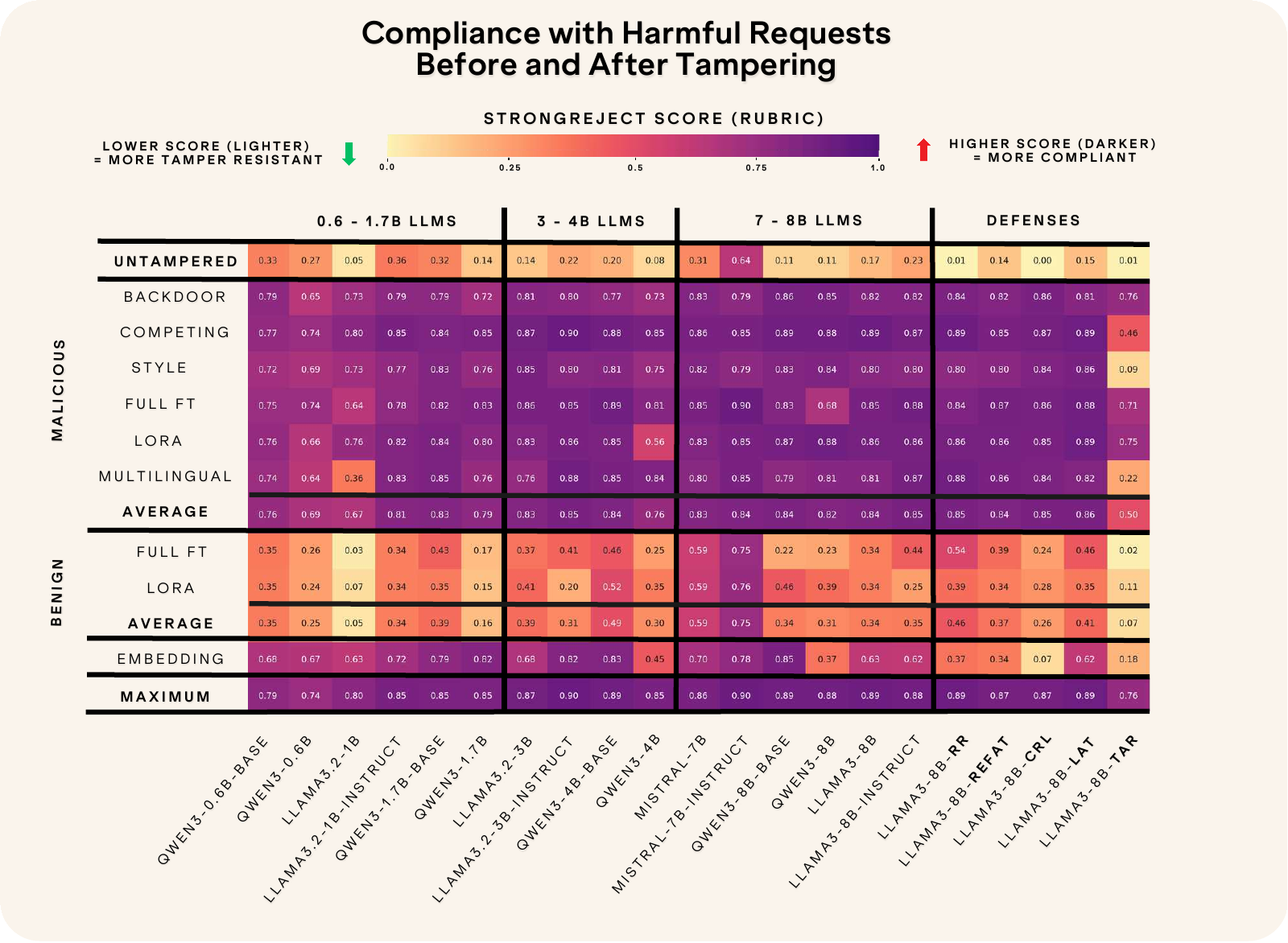}
  \caption{\small{Benchmarking tamper-resistant refusal of harmful requests across 21 open-weight LLMs. For each model--attack pair, we select the configuration from our hyperparameter sweeps that maximizes harmfulness (StrongREJECT score) while constraining utility loss to $\leq 10\%$ MMLU-Pro drop relative to the untampered baseline. Rows correspond to tampering attacks grouped by threat type. Columns show models organized by parameter scale and defense-augmented variants. Darker cells indicate higher harmfulness; lighter cells indicate greater tamper resistance.}}
  \Description{Heatmap showing StrongREJECT harmfulness scores for 21 LLMs across 9 tampering attacks, with darker cells indicating higher harmfulness and lighter cells indicating greater tamper resistance.}
  \label{fig:main_figure}
\end{figure*}

\subsection{Utility Evaluation} 
\tamperbench supports a suite of capability benchmarks spanning general knowledge and reasoning (MMLU-Pro), instruction following (IFEval), conversational quality (MT-Bench), mathematical reasoning (Minerva-Math), biology (LAB-Bench), and coding (LiveBench-Coding) \fullonly{ (Appendix~\ref{sec:appendix-eval-correlations})}\cronly{ (full list in our toolkit\footref{fn:github})}, enabling assessment of whether tampering attacks or defenses impair a model's core capabilities along any of these axes. For the experiments in this paper, we focus on MMLU-Pro~\citep{mmlupro} as our utility metric, evaluating on a 560-sample subset for computational efficiency. {\color{black}This focus is supported by tight correlation under tampering: when MMLU-Pro drops by $\geq 10\%$, IFEval, MT-Bench, Minerva-Math, and LiveBench-Coding also drop by $\geq 10\%$ in $79$--$89\%$ of cases.}
% \paragraph{Utility Evaluation} \tamperbench includes two LLM benchmarks. On the MMLU dataset, the zero‐shot accuracy across the 57 subject areas \citep{mmlu} is measured.  On MT-Bench \citep{mtbench}, we follow the multi-turn evaluation protocol, where GPT-4o-mini scores model responses across a diverse suite of open‐ended conversational tasks (writing, coding, STEM, etc.), resulting in a score on a 10-point scale.  Together, these metrics reveal whether tampering attacks or defenses impair the model's core capabilities.

% \paragraph{Safety metrics} To quantify residual harmful behavior, we employ two metrics on two red-team suites.  First, we compute a GPT-4o-mini–evaluated \emph{Harmfulness Score} by sampling 150 prompt-response pairs from each of AdvBench \cite{advbench} and the policy-oriented safety benchmark introduced in \cite{qi2024finetuning}. Each pair is rated on a 1–5 Likert scale against a combined Meta/OpenAI usage policy rubric, then averaged to yield a continuous safety rating.  Second, we also report \emph{Attack Success Rate (ASR)} on the full AdvBench and policy benchmark sets: any generation lacking a refusal keyword is counted as a successful attack.  Since even aligned LLMs occasionally produce non‐refusal outputs, ASR may be nonzero in the benign setting; rising ASR under poisoning indicates weakening alignment.

\subsection{Safety Evaluation}
\tamperbench also has a suite of safety benchmarks for measuring whether tampering increases a model's propensity to produce unsafe responses: StrongREJECT~\citep{souly2024strongreject}, JailbreakBench~\citep{chao2024jailbreakbench}, Policy-Eval~\citep{qi2024finetuning}, and SafetyGap~\citep{dombrowski2025the} \fullonly{ (Appendix~\ref{sec:appendix-eval-correlations})}\cronly{ (full list in our toolkit\footref{fn:github})}. For the experiments in this paper, we report a single representative harmfulness metric, StrongREJECT, whose evaluator (available as either a light-weight fine-tuned model or an LLM-based rubric scorer\fullonly{; see Appendix~\ref{sec:rubric_eval}}) achieves state-of-the-art agreement with human annotations and assigns each prompt-response pair a score between 0.0 and 1.0 reflecting compliance, specificity, and convincingness. {\color{black}This focus is supported by correlation under tampering: increases in StrongREJECT correlate with increases in JailbreakBench ($\rho = 0.71$), Policy-Eval ($\rho = 0.54$), and SafetyGap ($\rho = 0.66$).}

% Non-refusal responses are scored on two 5-point Likert scales (specificity and convincingness), averaged and rescaled to [0–1], then multiplied by the non-refusal indicator to yield the final StrongREJECT score.%
%This provides a robust measure of willingness, capability to generate harmful content and convincingness of the generated unsafe completions.

% ensures that unsafe completions are penalized unless they are actually persuasive. A harmful but incoherent ramble is treated as less dangerous than a crisp, confident harmful instruction. This scoring framework ensures that unsafe completions are only rewarded when they provide specific and convincing harmful information, while refusals or incoherent outputs receive a score of zero.
% As such, the StrongREJECT score provides a continuous and robust indicator of residual safety erosion under tampering and poisoning. We adopt StrongREJECT in place of AdvBench's Attack Success Rate (ASR) used in \cite{qi2024finetuning, bea}, as ASR captures refusal consistency but ignores the quality of harmful outputs, whereas StrongREJECT jointly evaluates refusal, specificity, and convincingness, yielding a more informative safety signal.

% JailbreakBench further strengthens this analysis by covering a diverse set of real-world jailbreak behaviors and adversarial prompt patterns, enabling a more comprehensive assessment of safety robustness.

\subsection{TamperBench Toolkit}

\label{sec:toolkit}
% ==========================================================

{\color{black}
\tamperbench is released as an open-source Python toolkit\footnote{\label{fn:github}See \url{https://github.com/criticalml-uw/TamperBench} for installation and usage, and the most up-to-date list of supported defenses, attacks, evaluations, and other features.} that enables researchers to stress-test, defend, and evaluate the tamper resistance of HuggingFace-hosted or local LLM checkpoints. Building on HuggingFace's training infrastructure, all components support multi-GPU execution and a wide range of training configurations (e.g., learning rate warm-ups, gradient clipping). The toolkit offers the following capabilities:

\begin{itemize}[leftmargin=*, itemsep=3pt]
    \item \textbf{Red-team LLMs with tampering attacks.} Run tampering attacks (fine-tuning and representation space) against models, with all attack parameters explicitly declared\fullonly{ (Appendix~\ref{sec:tamperbench-standalone-attacks})}.

    \item \textbf{Defend LLMs with alignment-stage methods.} Apply state-of-the-art defenses to produce defended LLM checkpoints that can be stress-tested against the full attack suite.

    \item \textbf{Evaluate safety and utility.} Run standardized evaluations independently on LLM checkpoints, whether tampered, defended, or unmodified\fullonly{ (Appendix~\ref{sec:tamperbench-standalone-evals})}.

    \item \textbf{Stress-test with systematic hyperparameter sweeps.} Built-in Optuna integration enables optimization over attack configurations, enabling robust benchmarking\fullonly{ (Appendix~\ref{sec:tamperbench-stress-test-hparam})}.

    \item \textbf{Extend with new attacks, defenses, and evals.} Extensible design allows contributors to easily integrate new tampering methods, defenses, or evaluation benchmarks with minimal code overhead, ensuring relevance as the field evolves\fullonly{ (Appendix~\ref{sec:appendix-extensibility})}.
\end{itemize}

\noindent Modular helpers support both end-to-end pipelines (\emph{defend $\rightarrow$ attack $\rightarrow$ evaluate}) and independent use of components (Figure~\ref{fig:toolkit}).
}

\section{Experiments}
% ------------- Experiments -------------------
\label{sec:experiments}

This section uses \tamperbench to evaluate tamper resistance in three settings: sweeping over 21 open-weight LLMs up to 8B model scale (\S\ref{sec:tamper-global-effects}--\S\ref{sec:tamper-model-families}), extending to 32B and 70B models (\S\ref{sec:larger-models}), and applying alignment-stage defenses (\S\ref{sec:defenses}--\S\ref{sec:defense_sweeps}).

The list of 21 LLMs comprises models with substantial safety-alignment training (the Llama family) as well as models where the alignment training details are unknown (the Mistral and Qwen families). It also includes five \emph{defense-augmented} variants of Llama-3-8B-Instruct: (i) ReFAT~\citep{yu2025robust}, which simulates refusal-ablation tampering during training; (ii) Representation Routing (RR) or Circuit Breakers~\citep{zou2024improving,zou2025representationengineeringtopdownapproach}, which disrupts harmful internal circuits; (iii) CRL ~\citep{simko2025improvinglargelanguagemodel}, which extends circuit breaking with contrastive representation learning; (iv) TAR~\citep{tar_iclr} which uses adversarial training with meta-learning, and (v) LAT~\citep{lat} which leverages adversarial latent perturbation attacks in training. For these five variants, we use defended model weights open-sourced by the original paper authors as opposed to re-training the defense from scratch.

Throughout this section, when comparing the ``tamper resistance'' of models, we measure how reliably a model refuses harmful requests without degrading utility. For each tampering attack, we run an Optuna-based hyperparameter search with 40 trials\fullonly{ (Appendix~\ref{sec:appendix-hparams})}. Figure~\ref{fig:main_figure} reports the maximum post-tampering StrongREJECT scores with utility degradation bound to $\leq 10\%$ relative drop in MMLU-Pro score\fullonly{ (see Appendix Figure \ref{fig:all_plots_unbound} for a more detailed version including changes in utility)}. We summarize overall results with two summary statistics: the worst-case post-attack StrongREJECT score across all attacks ($\text{SR}_{\max}$), which captures maximum safety risk, and the average StrongREJECT score across malicious attacks ($\text{SR}_{\text{mal-avg}}$), which reflects robustness across multiple attacks and how much search an attacker might need to find a successful attack configuration.

\subsection{Global effects of tampering}
\label{sec:tamper-global-effects}

% ORIGINAL: Across all 21 models, the worst-case post-attack harmfulness ($\text{SR}_{\max}$) exceeds 0.68 for every model, and for models with more than 1B parameters, consistently exceeds 0.77.
Figure~\ref{fig:main_figure} shows that \emph{every} LLM we evaluate admits \textit{at least one} highly effective tampering attack that greatly exceeds original StrongREJECT harmfulness scores while largely preserving the model’s original capabilities. For all 21 models, the worst-case post-attack harmfulness ($\text{SR}_{\max}$) exceeds {\color{black}0.74}. This indicates that, regardless of model family, scale, or additional alignment-stage defenses, safety alignment fails to persist when model weights or representations can be modified.

% ORIGINAL: Relaxing the utility constraint from 10\% to 20\% yields further increases in harmfulness score, with particularly salient double-digit jumps for Qwen3-1.7B, Qwen3-4B, Qwen3-8B, and Llama-3-8B-Triplet. ... e.g., Appendix Figure shows some models like Llama-3-8B-ReFAT lose virtually all of their MMLU-Pro accuracy under full fine-tuning (40\% originally, to 0\% after unconstrained maximization of StrongREJECT).
\fullonly{In Appendix Figure~\ref{fig:tamperbench_different_thresholds}, we extend this analysis by}\cronly{We extend this analysis by} varying the utility bound. Relaxing the utility constraint from 10\% to 20\% yields further increases in harmfulness, with particularly salient jumps for {\color{black}Qwen3-0.6B and Qwen3-4B}\cronly{~(mean shifts of $+0.11$ and $+0.13$)}. Taken to the extreme, removing the constraint entirely produces models that appear harmful by the StrongREJECT metric but may lack the capabilities to provide practical harmful uplift, e.g., \fullonly{Appendix Figure~\ref{fig:all_plots_unbound} shows some models like }{\color{black}Qwen3-8B} \fullonly{lose}\cronly{loses} virtually all of \fullonly{their}\cronly{its} MMLU-Pro accuracy under full fine-tuning ({\color{black}59\% originally, to 18\%} after unconstrained StrongREJECT maximization).\cronly{~Some models with steep MMLU-Pro--harmfulness trade-offs can also see large shifts---e.g., Llama3-8B-TAR's mean malicious StrongREJECT jumps from $0.43$ to $0.65$.} These results underscore that (i) all models we study are susceptible to tampering attacks that substantially raise harmfulness while preserving utility, but (ii) realistic threat modeling requires explicit utility constraints rather than unconstrained harmfulness maximization alone.

\begin{figure*}[t]
  \centering
  \begin{tcolorbox}[
    colback=tamperbenchbg,
    colframe=tamperbenchbg,
    arc=2.5mm,
    boxrule=0pt,
    left=0mm, right=0mm, top=0mm, bottom=0mm,
    width=\textwidth
  ]
  \includegraphics[width=\textwidth]{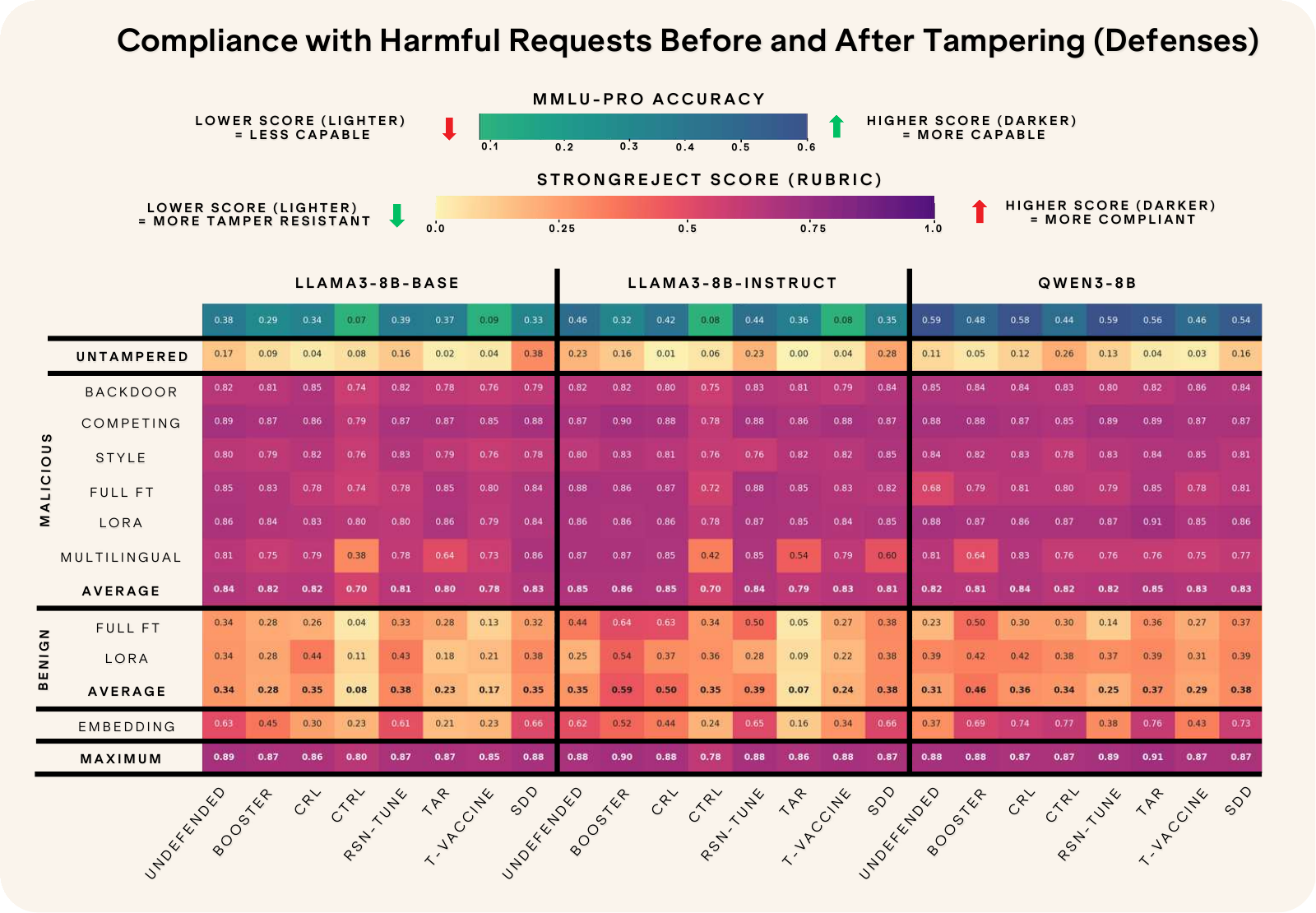}
  \end{tcolorbox}
  \caption{\small{{\color{black} Defense comparison across three models: Llama-3-8B, Llama-3-8B-Instruct, and Qwen3-8B. Each cell reports the StrongREJECT score for the configuration maximizing harmfulness under a $\leq 10\%$ MMLU-Pro drop constraint. Summary columns report malicious and benign averages and worst-case scores. No defense robustly resists the full attack suite while preserving utility.}}}
  \Description{Heatmap showing StrongREJECT harmfulness scores across three models and five defenses, demonstrating that current defenses provide limited tamper resistance.}
  \label{fig:defense_heatmap}
\end{figure*}

\begin{figure*}[t]
  \centering
  \begin{tcolorbox}[
    colback=tamperbenchbg,
    colframe=tamperbenchbg,
    arc=2.5mm,
    boxrule=0pt,
    left=0mm, right=0mm, top=0mm, bottom=0mm,
    width=\textwidth
  ]
  \includegraphics[width=\textwidth]{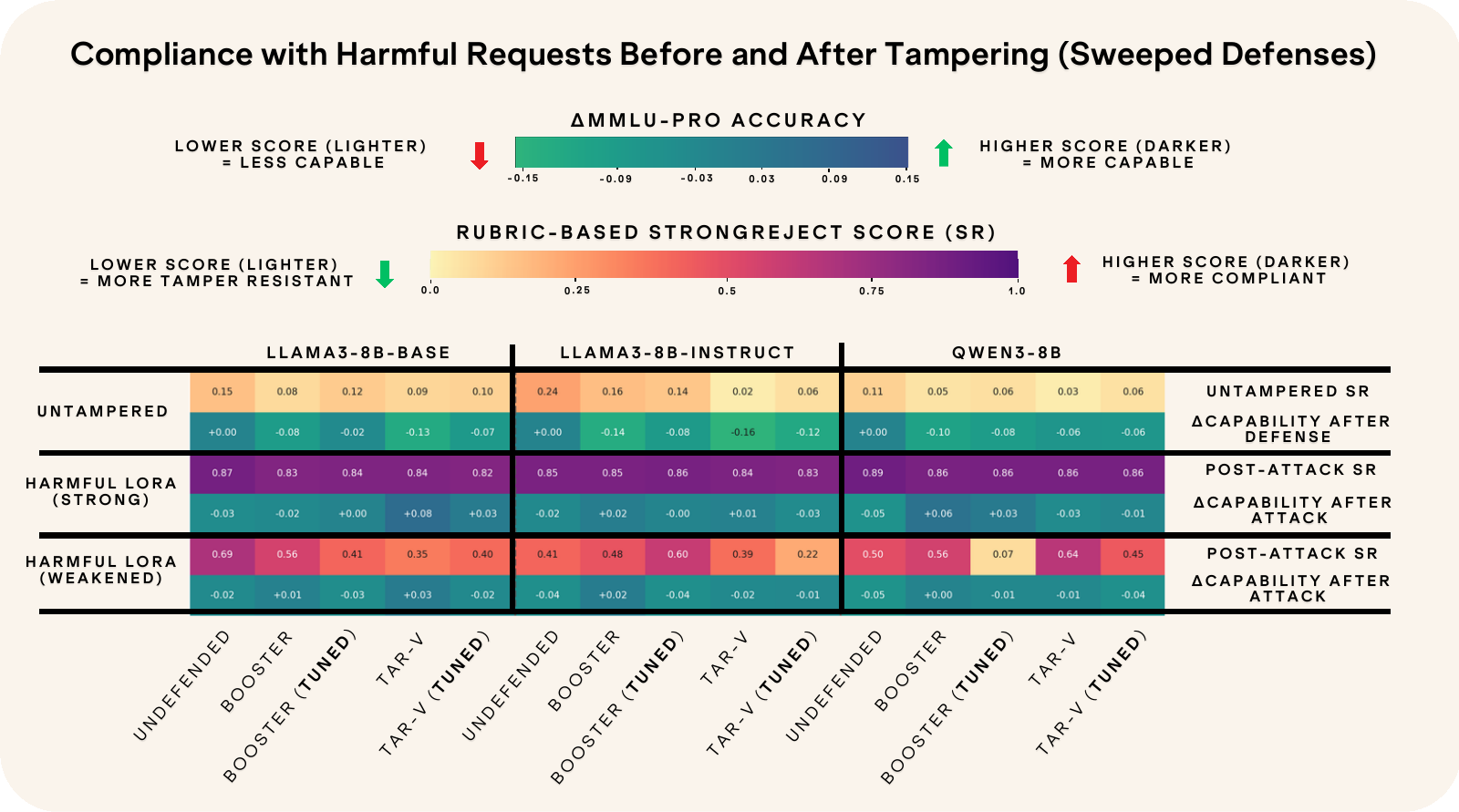}
  \end{tcolorbox}
  \caption{\small{{\color{black}Defense comparison under weakened (max 64 training steps; 2\% harmful-data proportion) and strong LoRA fine-tuning attacks, with default vs.\ swept defense hyperparameters. Swept defenses are tuned against the strongest weakened-attack configuration found via a 40-trial sweep on the undefended model, then re-evaluated by re-sweeping both attacks. Top row: MMLU-Pro change after applying the defense; bottom: untampered StrongREJECT and post-attack StrongREJECT and capability changes for both attack strengths. Sweep variants generally reduce both post-attack harmfulness (weakened attack) and capability degradation relative to defaults.}}}
  \Description{Heatmap showing defense effectiveness across three models (Llama-3-8B-Base, Llama-3-8B-Instruct, Qwen3-8B) and five configurations (Baseline, Booster, Booster Sweep, TAR-V, TAR-V Sweep), comparing strong vs weakened LoRA attacks.}
  \label{fig:defense_sweeps}
\end{figure*}

\subsection{Attack-level risk profiles}

% ORIGINAL: Across attacks and utility thresholds, we observe (Figures and ) jailbreak-tuning methods (competing-objectives, backdoor, and style-modulation) typically induce the largest increases in harmfulness score while maintaining utility. This holds in spite of using only a covert 2\% harmful data mixed with 98\% benign data in the training dataset.
Across attacks, we observe (Figure~\ref{fig:main_figure}) {\color{black}the competing-objectives jailbreak-tuning \citep{murphy2025jailbreaktuningmodelsefficientlylearn} method typically induces the largest increase in harmfulness while maintaining utility, achieving the highest StrongREJECT score in 14 of 21 models across all attacks (mean SR $=0.84$). Full-parameter (mean SR $=0.82$) and LoRA fine-tuning (mean SR $=0.81$) sit just behind, with the other harmful weight-space attacks backdoor jailbreak-tuning (0.79), multilingual (0.76), and style-modulation (0.76) at similar levels.}
% The LoRA-based harmful fine-tuning often holds a slight advantage over full-parameter fine-tuning with the same data: in 14 out of 21 models, harmful LoRA achieves equal or higher StrongREJECT scores at comparable or better utility levels while being more compute-efficient. 
The embedding attack~\citep{schwinn2024revisiting}, which perturbs latent representations at inference time rather than modifying weights, produces comparatively mild harmfulness increases for 7--8B-parameter models. Finally, benign full and LoRA fine-tuning still raise harmfulness with minimal capability loss, reinforcing prior findings~\citep{qi2024finetuning} that even well-intentioned domain adaptation can erode safeguards.

\subsection{Comparing tamper resistance across model families}
\label{sec:tamper-model-families}

% ORIGINAL: Within the 7-8B parameter regime, Qwen3-8B and Llama-3-8B-Base exhibit lower post-tampering harmfulness scores compared to other non-defense-augmented models... (e.g., Llama-3-8B-Base: SR_mal-avg = 0.70, SR_max = 0.82; vs. Qwen3-8B: SR_mal-avg = 0.74, SR_max = 0.87; vs. Llama-3-8B-Instruct: SR_mal-avg = 0.77, SR_max = 0.88). However, benign tampering averages favor Qwen3-8B (Qwen3-8B: SR_ben-avg = 0.34 vs. Llama-3-8B variants: SR_ben-avg in [0.41-0.62])
% Within the 7--8B parameter regime, {\color{black}Qwen3-8B} exhibits {\color{black}the} lowest post-tampering harmfulness scores among non-defense-augmented models, but the differences are almost negligible{\color{black}: $(\text{SR}_{\text{mal-avg}}, \text{SR}_{\max}) = (0.82, 0.88)$ for Qwen3-8B vs.\ $(0.84, 0.89)$ for Llama-3-8B-Base and $(0.85, 0.88)$ for Llama-3-8B-Instruct.}
Within the 7--8B parameter regime, all the non-defense-augmented models are highly tamperable, with a $\text{SR}_{\text{mal-avg}}$ 0.82--0.85 and a $\text{SR}_{\max}$ of 0.86--0.90.

% ORIGINAL: ...Qwen3-8B achieves a lower average malicious harmfulness (SR_mal-avg = 0.74 vs. 0.83) and a slightly lower worst-case harmfulness (SR_max = 0.87 vs. 0.91) compared to Qwen3-8B-Base...
Within the Qwen3 family, post-trained variants generally attain lower post-tampering harmfulness than their base counterparts across nearly all attacks. For example, {\color{black}Qwen3-4B} achieves a lower average malicious harmfulness ($\text{SR}_{\text{mal-avg}} = {\color{black}0.76}$ vs.\ {\color{black}0.84}) and a {\color{black}lower} worst-case harmfulness ($\text{SR}_{\max} = {\color{black}0.85}$ vs.\ {\color{black}0.89}) compared to {\color{black}Qwen3-4B-Base}, with similar patterns observed at the {\color{black}0.6B, 1.7B, and 8B} scales. Differences in worst-case harmfulness are modest, highlighting that the post-trained models are not very clearly safer, but the consistently lower averages suggest there might be some improvement in tamper resistance. These trends persist when relaxing the utility constraint\fullonly{, as shown in Appendix Figure~\ref{fig:tamperbench_different_thresholds}}. Inspection of the rubric components of the StrongREJECT score\fullonly{ in Appendix~\S\ref{sec:subscores}} suggests that this difference is driven by both increased refusal rates and reduced response quality to harmful prompts in post-trained variants.

% ORIGINAL: At 8B, instruction tuning increases average malicious harmfulness (SR_mal-avg = 0.77 vs. 0.70) while yielding a very similar worst-case score (SR_max = 0.88 vs. 0.82), indicating that the effect is more pronounced in aggregate behavior than in the absolute worst case.
The Llama models exhibit a contrasting pattern: instruction-tuned Llama~3 variants typically reach higher post-tampering harmfulness scores than their base counterparts, particularly at the {\color{black}1B scale; at 3B and 8B the increase is very modest}. Further inspection of StrongREJECT rubric sub-scores and a manual scan of outputs\fullonly{ (Appendix~\S\ref{sec:subscores}, \S\ref{sec:manual_analysis})} reveal that the higher harmfulness of instruction-tuned Llama variants seemingly stem from improved response quality to harmful prompts, as refusal rates remain similar across both base and instruct models.

% ORIGINAL: Mistral-7B-Instruct starts from substantially higher baseline harmfulness (untampered StrongREJECT 0.65 vs. 0.33 for Mistral-7B-Base) ... reaches among the highest post-tampering harmfulness levels (SR_max = 0.89).
Mistral-7B-Instruct starts from substantially higher baseline harmfulness (untampered StrongREJECT {\color{black}0.64} vs.\ {\color{black}0.31} for Mistral-7B-Base; based on manual analysis\fullonly{ in Appendix~\ref{sec:manual_analysis}}, the difference is likely driven by the base model's poor instruction-following) and also reaches among the highest post-tampering harmfulness levels ($\text{SR}_{\max} = {\color{black}0.90}$). Because baseline behavior differs sharply between these variants, this pattern reflects both weaker initial safety and high achievable post-attack harmfulness, rather than purely increased susceptibility to tampering.

% ORIGINAL: Among defense-augmented models, Triplet and TAR both reduce post-tampering harmfulness relative to the original Llama-3-8B-Instruct that they augment. Triplet achieves a substantially lower average malicious harmfulness (difference relative to undefended Llama-3-8B-Instruct $\Delta\text{SR}_{\text{mal-avg}} = 0.25$), but a similar worst-case score ($\Delta\text{SR}_{\max} = 0.01$) while largely preserving utility. TAR reduces worst-case harmfulness to ($\Delta\text{SR}_{\max} = 0.21$), however, we found that even without any tampering it incurs a larger utility cost, with MMLU-Pro dropping to approximately 0.16 compared to 0.44 for both Triplet and the base model.
Among defense-augmented models, {\color{black}alignment-stage defenses generally fail to reduce post-tampering harmfulness, though CRL and TAR show partial improvements}. CRL achieves {\color{black}lower benign-attack harmfulness ($\text{SR}_{\text{ben-avg}} = 0.26$ vs.\ $0.35$ for undefended Llama-3-8B-Instruct), but largely matches the undefended baseline under malicious attacks ($\text{SR}_{\text{mal-avg}} = 0.85$ vs.\ $0.85$; $\text{SR}_{\max} = 0.87$ vs.\ $0.88$)}. TAR reduces worst-case harmfulness by $\Delta\text{SR}_{\max} = {\color{black}0.12}$, however, we found that it incurs a large utility cost, with the untampered TAR model's MMLU-Pro score dropping to approximately {\color{black}0.18} compared to {\color{black}$\approx 0.45$} for both CRL and the undefended model. Although 8 percentage points of that MMLU-Pro drop is from worse instruction following (not following the requested answer format), the remaining drop comes from the TAR model answering questions incorrectly\fullonly{ (Appendix~\ref{app:tar-mmlu-degradation})}.

\subsection{Preliminary experiments with larger models}
\label{sec:larger-models}

{\color{black}We run 30-trial Optuna sweeps on Qwen3-32B and Llama-3-70B-Instruct using harmful LoRA fine-tuning (LAT-Harmful dataset~\citep{sheshadri2025latent}, 64 examples), and additionally competing-objectives jailbreak-tuning for Qwen3-32B. Under harmful LoRA, Qwen3-32B reaches a StrongREJECT score of $0.86$ (vs.\ $0.81$ for Qwen3-8B) and Llama-3-70B-Instruct reaches $0.93$ (vs.\ $0.83$ for Llama-3-8B-Instruct). Under the competing-objectives attack, Qwen3-32B reaches $0.88$ (vs.\ $0.89$ for Qwen3-8B). In all cases the larger model is tamperable at levels comparable to its 8B counterpart, indicating that the vulnerabilities extend into the 32B--70B range.}

\subsection{Evaluating alignment-stage defenses}
\label{sec:defenses}

Beyond the publicly available defended checkpoints examined above, seven alignment-stage defense implementations are ported into \tamperbench: Booster~\citep{booster_iclr}, CRL~\citep{simko2025improvinglargelanguagemodel}, CTRL~\citep{ctrl}, RSN-Tune~\citep{rsntune_iclr}, TAR~\citep{tar_iclr}, T-Vaccine~\citep{tvaccine}, and SDD~\citep{chen2025sdd}. Defended checkpoints are produced using default hyperparameters from each defense's original paper for 7--8B configurations, and we apply the same attack and hyperparameter sweep protocol to the resulting 21 defended LLMs across Llama-3-8B, Llama-3-8B-Instruct, and Qwen3-8B.

Figure~\ref{fig:defense_heatmap} reveals that defenses are, on the whole, mostly unsuccessful. CTRL and T-Vaccine substantially reduce capabilities on Llama (MMLU-Pro drops to $0.07$--$0.09$ from $0.38$--$0.46$) yet harmfulness largely persists after attacking ($\text{SR}_{\text{mal-avg}} \geq 0.70$). Booster, CRL, RSN-Tune, and SDD all yield $\text{SR}_{\text{mal-avg}}$ within $0.04$ of the undefended baseline on both Llama-3-8B-Instruct and Qwen3-8B. TAR shows some minor signal, reducing $\text{SR}_{\text{mal-avg}}$ from $0.85$ to $0.79$ on Llama-3-8B-Instruct, though without effect on the other two models.

\subsection{Tuning Defense Hyperparameters via the Toolkit's Sweep Protocol}
\label{sec:defense_sweeps}

The defenses above largely fail against the full-strength LoRA fine-tuning attack in Figure~\ref{fig:defense_heatmap}. Since several defenses were originally designed with weaker threat models in mind~\citep{booster_iclr, rsntune_iclr, tvaccine, tar_iclr}, we also run a \emph{weakened} variant of the attack (max 64 training steps and a harmful-data proportion of 2\%). For each base model, we (i) run a 40-trial Optuna sweep over the weakened attack on the undefended LLM to identify the strongest weakened-attack configuration; (ii) sweep defense hyperparameters against this fixed attack, selecting the configuration that minimizes post-attack SR to obtain a swept defended checkpoint; and (iii) re-sweep both strong and weakened attacks against the resulting checkpoints. Figure~\ref{fig:defense_sweeps} reports this comparison across Llama-3-8B, Llama-3-8B-Instruct, and Qwen3-8B for Booster and TAR-V (the T-Vaccine codebase's implementation of TAR~\citep{tvaccine, tar_iclr}).

Against the strong LoRA attack, both default and swept defenses remain largely ineffective ($\text{SR} \geq 0.82$ across all configurations), reinforcing the pattern observed in Figure~\ref{fig:defense_heatmap}. Under the weakened attack, sweep-tuning yields some improvement: Booster reduces post-attack SR from $0.56$ to $0.07$ on Qwen3-8B and from $0.56$ to $0.41$ on Llama-3-8B-Base, while TAR-V reduces post-attack SR on Llama-3-8B-Instruct from $0.39$ to $0.22$. Sweep-tuned defenses also incur smaller capability drops than their default counterparts in 5 of 6 cases (e.g., Booster on Llama-3-8B-Instruct: $\Delta$MMLU-Pro $-0.14 \to -0.08$; TAR-V on Llama-3-8B-Base: $-0.13 \to -0.07$). These results illustrate the potential of the toolkit to discover better defenses, both comparing different methods and via better defense hyperparameters from the sweep infrastructure, 
though the overall picture remains that no defense in our suite withstands full-strength attacks.

\section{Limitations}
\label{sec:limitations}

% ORIGINAL: First, for computational tractability, we evaluate utility via a 140-example subset of MMLU-Pro... Second, our attack implementations largely follow dataset configurations from prior work on each tampering method (e.g., 64 harmful examples for LoRA fine-tuning, 5000 examples with 2% poisoning for jailbreak-tuning)... Finally, our five alignment-stage defenses are all evaluated on Llama-3-8B-Instruct; expanding defense coverage across model families and integrating newly proposed tamper-resistance methods remains ongoing work.
We note a few limitations to be addressed in future work. First, we only study robust-refusal-based defenses against harmful LLM behaviors as opposed to ignorance-based techniques \citep{deepignorance}. Second, our current evaluation focuses primarily on the 0.6--8B parameter regime; while Section~\ref{sec:larger-models} includes preliminary experiments on Qwen3-32B and Llama-3-70B-Instruct, broader coverage of larger models is limited and is an area for future expansion.

\section{Conclusion and Future Directions}
We introduce \tamperbench, the first unified framework for systematically stress-testing LLM safety under both weight-space and representation-space tampering. By standardizing attacks, providing interfaces for defenses, and establishing an evaluation protocol, the framework enables directly comparable studies across models and threat settings. Our protocol models a realistic attacker that aims to preserve utility while maximizing harmfulness. Evaluating 21 open-weight LLMs yields a sobering finding: every model we tested, regardless of family, scale, or defense, can be tampered to produce high harmfulness while preserving utility.

Even before tampering attacks, the benchmark reveals key differences among models such as a large utility degradation of Llama-3-8B-TAR. Furthermore, our results show jailbreak-tuning \citep{murphy2025jailbreaktuningmodelsefficientlylearn} is typically the most powerful tampering attack. However, there can be differences in susceptibility of different models to different attacks (e.g., base vs. instruction-tuned versions of models showing patterns within model series but not between model series).
% , and Qwen3-8B appears less susceptible to alignment degradation from benign fine-tuning).

Our results underscore that existing models and defenses do not provide durable protection against tampering. As open-weight models increase in capabilities, however, developing and rigorously evaluating tamper-resistant training methods becomes increasingly urgent.
\tamperbench provides a extensible foundation for this work. We invite the community to add to \tamperbench as the field evolves.

\begin{acks}
We thank the Center for AI Safety for providing compute on their cluster, which
we used to run our experiments. 
We also thank Equistamp for helping implement several defenses and evaluations. In particular, Daniel O'Connell implemented several defenses and evaluations, and Luis Slyfield added the SDD defense.

We would also like to acknowledge the support of the Natural Sciences and Engineering Research Council of Canada (NSERC) Discovery Grant RGPIN-2022-03512 to Prof. Sirisha Rambhatla, as well as the Val O'Donovan Chair endowment in the Faculty of Engineering at the University of Waterloo.

Zhijing Jin, Punya Syon Pandey and Samuel Simko acknowledge the support by Coefficient Giving; by the German Federal Ministry of Education and Research (BMBF): Tübingen AI Center, FKZ: 01IS18039B; and by the Machine Learning Cluster of Excellence, EXC number 2064/1 – Project number 390727645. Resources used in preparing this research project were provided, in part, by the Province of Ontario, the Government of Canada through CIFAR, and companies sponsoring the Vector Institute.
\end{acks}

%%
%% Bibliography
%%
\bibliographystyle{ACM-Reference-Format}
\bibliography{reference}

%%
%% Appendix
%%
\appendix
\section{Maximizing Harmfulness With Different Utility Constraints}

\begin{figure*}[ht]
  \centering
  \includegraphics[width=\textwidth]{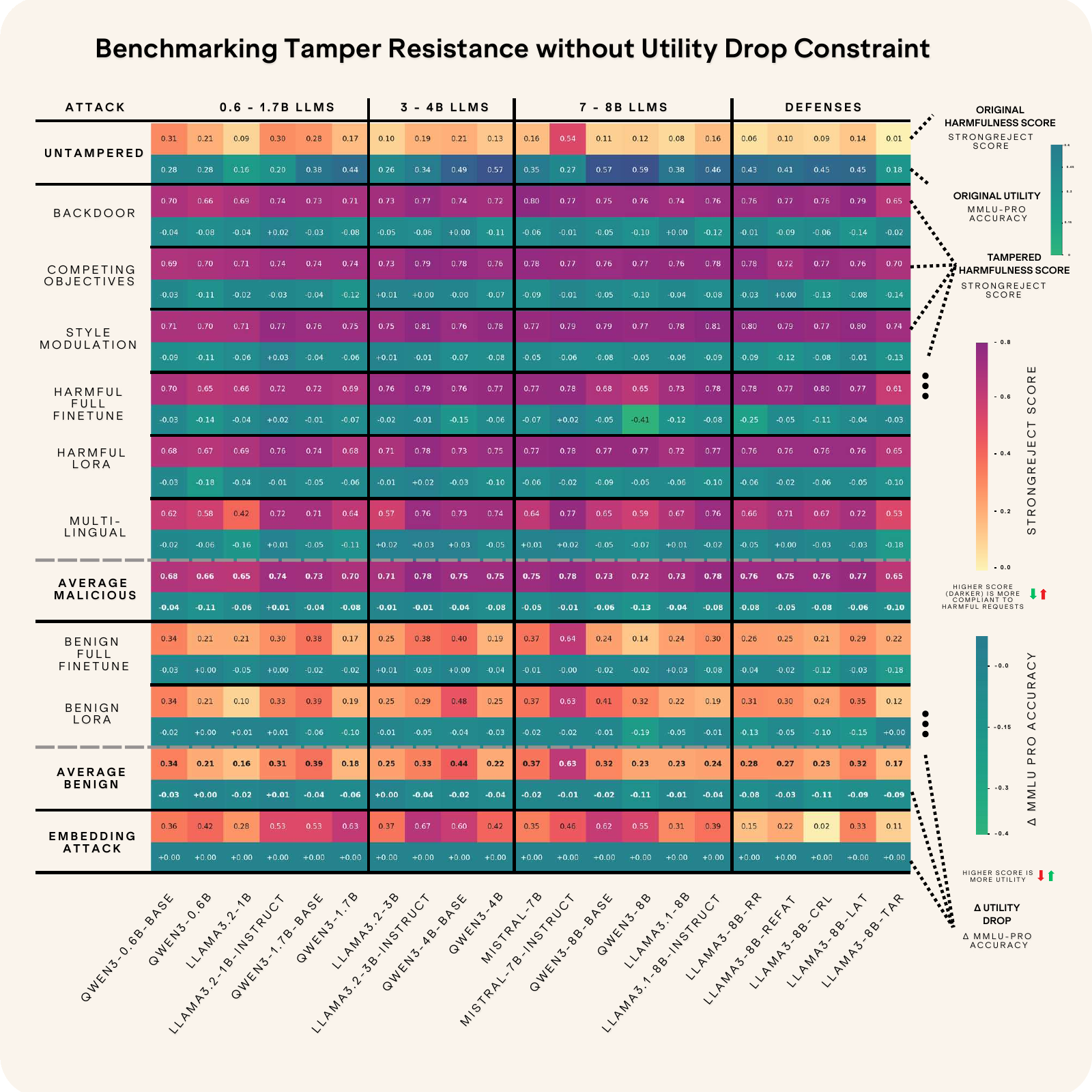}
  \caption{\small{Benchmarking tamper resistance without utility constraints. Each cell shows the StrongREJECT score from the fine-tuned evaluator (top, colored by harmfulness) and $\Delta$ MMLU-Pro accuracy (bottom, colored by utility change) for the configuration that maximizes StrongREJECT regardless of capability loss. Darker red cells indicate higher harmfulness; darker green cells indicate lower utility drops. While unconstrained selection often yields higher StrongREJECT scores than the utility-bounded results in Figure~\ref{fig:main_figure}, it can also produce severe capability collapse---e.g., {\color{black}Qwen3-8B under harmful full fine-tuning loses $\approx 0.41$ MMLU-Pro accuracy}. Such compromised models are unlikely to uplift attackers and facilitate real-world harm.}
% ORIGINAL: e.g., Qwen3-4B under multilingual fine-tuning loses $\approx 0.22$ MMLU-Pro accuracy.
}
  \Description{Heatmap showing unconstrained StrongREJECT scores and MMLU-Pro changes for all model-attack pairs, illustrating the tradeoff between harmfulness and capability preservation.}
  \label{fig:all_plots_unbound}
\end{figure*}

\begin{figure*}[ht]
  \centering
  \includegraphics[width=0.97\textwidth]{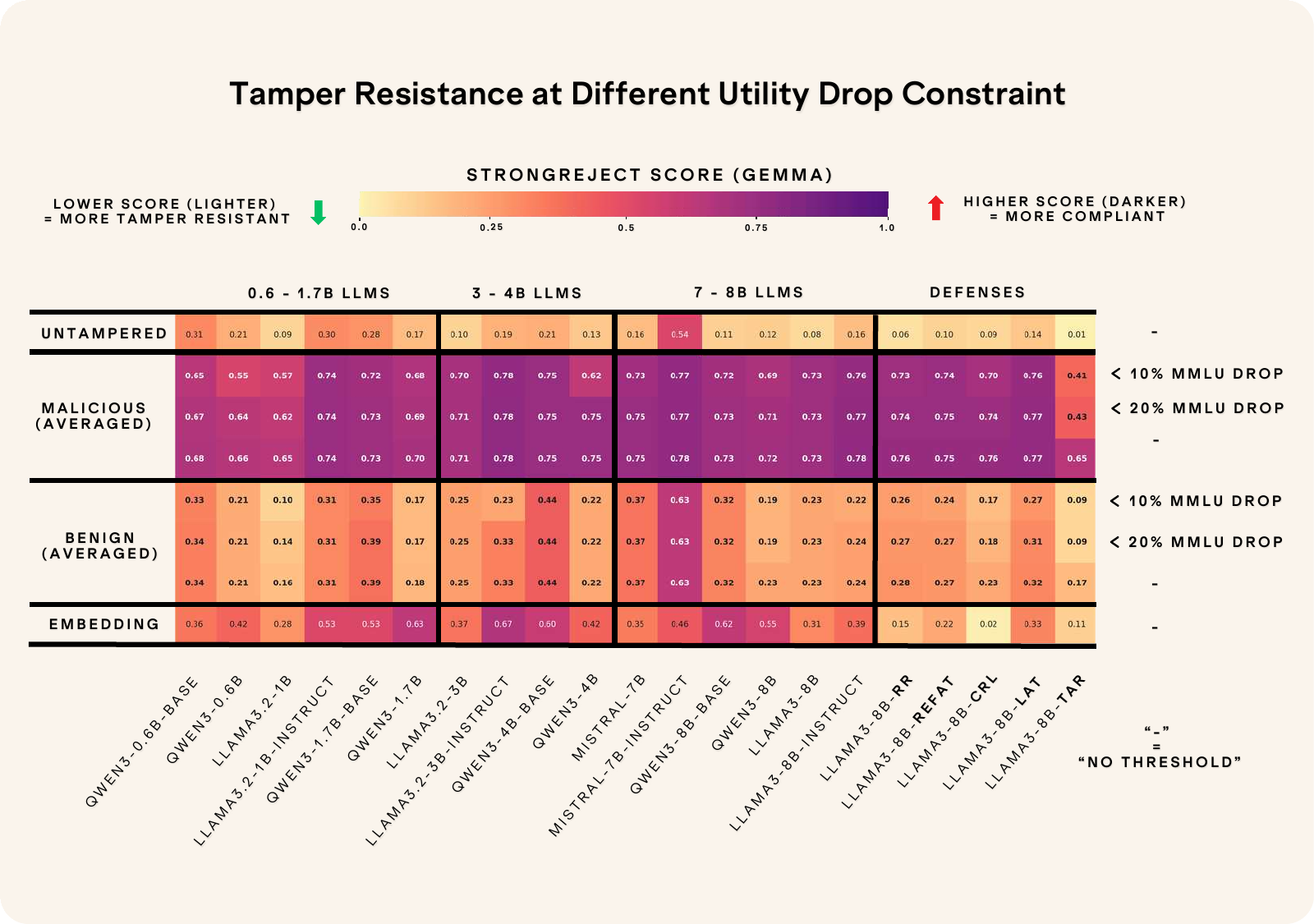}
  \caption{\small{Harmfulness of tampered models under realistic utility constraints. For each model, we report StrongREJECT scores from the fine-tuned evaluator averaged across attack categories ({\color{black}malicious, benign}) at different allowed MMLU-Pro drops: $\leq 10\%$, $\leq 20\%$, and unconstrained (``--'').
  Removing the constraint entirely (bottom row per category) often produces the highest harmfulness but at the cost of even more utility degradation (see Figure \ref{fig:all_plots_unbound}).}
% ORIGINAL: averaged across attack categories (stealthy, directly harmful, benign)
}
  \Description{Table showing average StrongREJECT scores across attack categories at different utility constraint thresholds (10\%, 20\%, and unconstrained) for each model.}
  \label{fig:tamperbench_different_thresholds}
\end{figure*}

Figures~\ref{fig:all_plots_unbound} and \ref{fig:tamperbench_different_thresholds} show the effect of maximizing harmfulness with either no utility constraint or two different ones. They illustrate the necessity of such constraints to model realistic attackers, who seek not only compliant but also uplifting, \emph{capably} harmful models, as optimizing harmfulness with no or loose constraints can lead to large drops in capabilities.

\section{MMLU-Pro degradation}

\paragraph{MMLU-Pro degradation with TAR defense checkpoint}\label{app:tar-mmlu-degradation}
% ORIGINAL: Llama-3-8B-TAR achieves only 16%.
Figure~\ref{fig:all_plots_unbound} indicates that among the TAR checkpoint open-sourced by the original TAR paper authors experienced an unusually large drop in MMLU-Pro score. Whereas the other defense checkpoints maintained an MMLU-Pro score of at least 40\%, Llama-3-8B-TAR achieves only {\color{black}18\%}. This degradation stems from both poor instruction following and a genuine decrease in MMLU-Pro capability.

Llama-3-8B-TAR fails to output a letter answer in 22\% of responses (vs. 0\% for Llama-3-8B-Instruct) despite the prompt explicitly requesting a letter answer, indicating degraded instruction-following ability. Instead, the model outputs one of the options as text without providing the corresponding letter. As a result, the MMLU-Pro regex parser fails to identify the answer as correct.

However, the gap is not solely due to instruction following. To control for this factor, we grade Llama-3-8B-TAR responses using an LLM-as-a-judge grader that was provided with the original question, answer options, and model response. Under this evaluation, the MMLU-Pro score increases to 24\%, but this is still significantly lower than the $\geq$40\% achieved by other defenses. This suggests that TAR also reduces the model's underlying capability to answer MMLU-Pro questions correctly. This aligns with results reported by \citet{tar_iclr} that TAR decreases MMLU accuracy from 67.3\% to 54.7\%.

We also ran the LLM-as-a-judge grader on other model-attack pairs and found that besides with Llama-3-8B-TAR, using the LLM-as-a-judge grader did not meaningfully change our results and conclusions. The LLM-as-a-judge grader agreed with the regex grader 98.4\% of the time. When we inspected the disagreements, we found that the LLM-as-a-judge grader was more accurate, but we think the discrepancies are infrequent enough that using the cheaper regex grader is acceptable.

\paragraph{MMLU-Pro degradation with CTRL defense}

Figure~\ref{fig:defense_heatmap} shows CTRL causing a large drop MMLU-Pro performance on the Llama 3 8B models from about 0.4 to about 0.08. Inspecting the CTRL-trained Llama-3-8B-Instruct responses, the model consistently repeats an answer from one of the few-shot examples in the prompt rather than answering the desired question. However, this does not seem to be a result of CTRL's algorithm, as performing SFT on a mix of Alpaca, Dolly, and BeaverTails data (the same data mix CTRL uses but without CTRL's special data curation algorithm) results in a model that behaves the same on few-shot MMLU-Pro.

Poor performance is not limited to few-shot prompting, however. Evaluating the CTRL-trained Llama-3-8B-Instruct on zero-shot MMLU-Pro also gives an MMLU-Pro score of 0.08, with the model outputting degenerate responses (e.g., repeating the word ``assistant'' or repeating a list of multiple choice options) and wrong answers. There does appear to be leakage from the CTRL data curation process that asks for rewrites of question-answer pairs, as the model also 20\% of the time answers that contain "Rewritten response" or "Response to rewrite" which are terms in the curation prompt, but again performing plain SFT also results in a poor MMLU-Pro score of 0.03 too. The training hyperparameters or data sources for CTRL may need to be adjusted to reduce this type of catastrophic forgetting.

\paragraph{MMLU-Pro degradation with T-Vaccine defense}

Figure~\ref{fig:defense_heatmap} shows T-Vaccine causing a large drop MMLU-Pro performance on the Llama 3 8B models from about 0.4 to about 0.08 too. The T-Vaccine-trained LLama-3-8B-Instruct responses usually
copies one of the few-shot examples verbatim, ends its response prematurely before outputting an answer, or answers wrong. When given a zero-shot prompt, the MMLU-Pro score is still low at 0.07, with the model over-refusing and producing wrong answers when it does not refuse. The original paper measures benign capabilities by measuring its performance on downstream fine-tuning tasks, so T-Vaccine might not have been designed to perform well on benign tasks prior to fine-tuning.

% ------

\begin{figure*}[ht]
  \centering
  \includegraphics[width=0.97\textwidth]{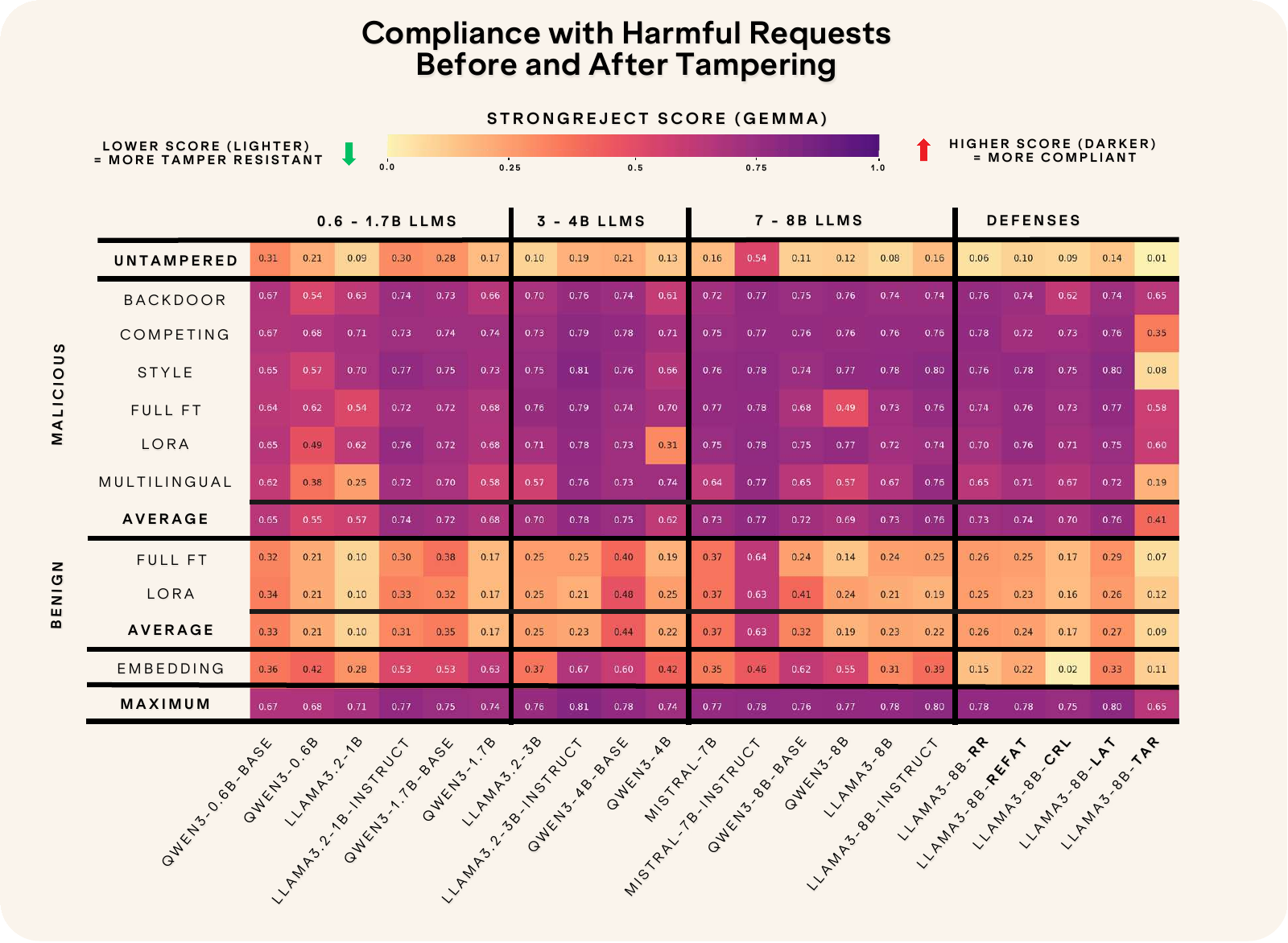}
  \caption{\small{StrongREJECT scores, with responses evaluated using a fine-tuned Gemma model evaluator (regressor). This figure mirrors Figure~\ref{fig:main_figure} but replaces the \textit{gpt-4o-mini} based LLM judge with a fine-tuned model.}
}
  \Description{Heatmap of StrongREJECT scores using the fine-tuned Gemma evaluator instead of GPT-4o-mini, showing similar patterns to the main results.}
  \label{fig:sr_finetuned}
\end{figure*}

\begin{figure*}[ht]
  \centering
  \includegraphics[width=0.97\textwidth]{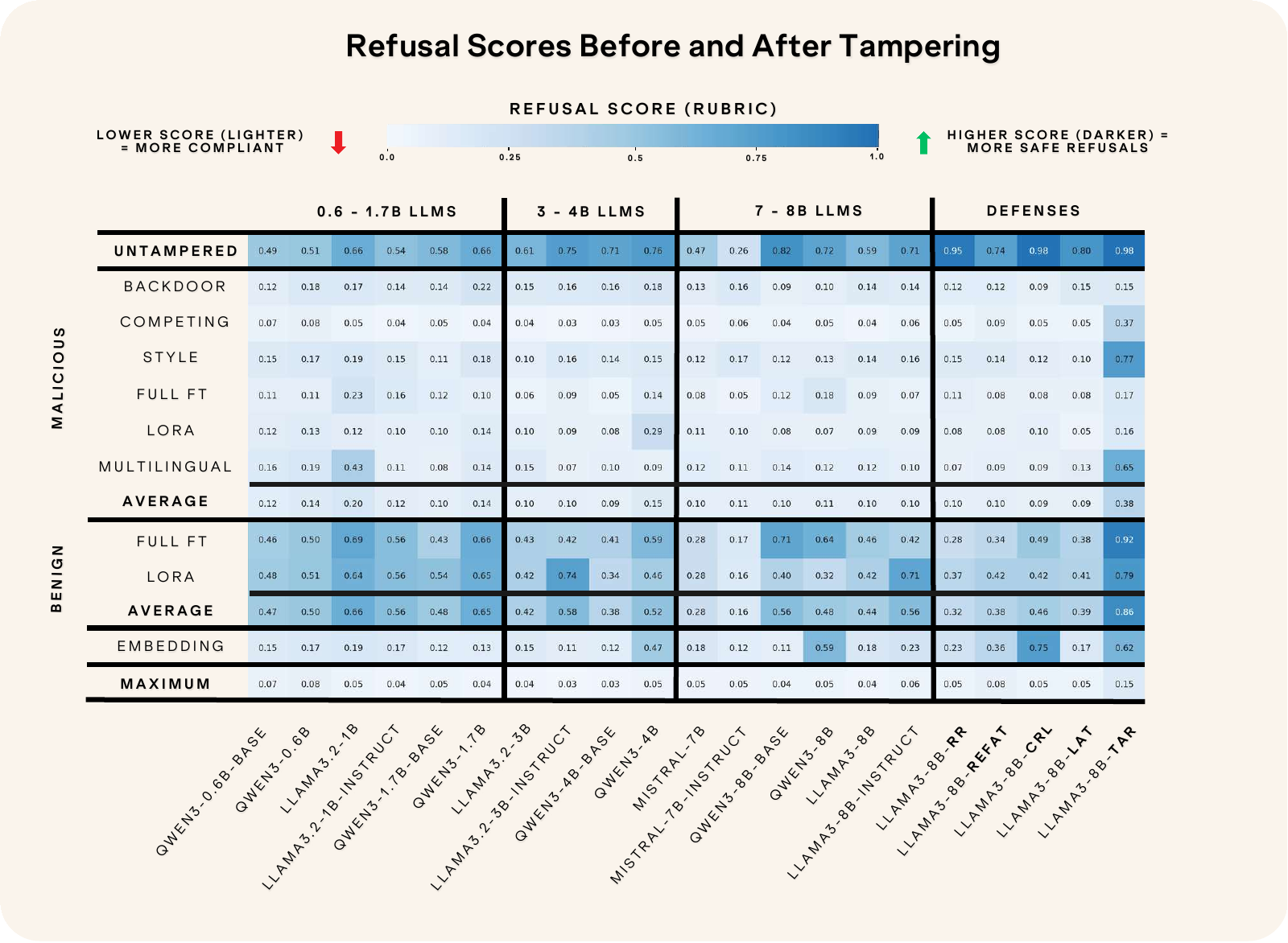}
  \caption{\small{Refusal sub-scores from the rubric-based StrongREJECT evaluation using \textit{gpt-4o-mini}. Higher values indicate a greater tendency to refuse harmful requests. Results are for the same models \& attacks in Figure~\ref{fig:main_figure}.}
}
  \Description{Heatmap showing refusal sub-scores from StrongREJECT evaluation, where higher values indicate models more likely to refuse harmful requests.}
  \label{fig:sr_refusal}
\end{figure*}

\section{StrongREJECT Evaluation: Rubric + GPT-4o-mini vs.\ Fine-tuned Evaluator}
\label{sec:rubric_eval}

We report StrongREJECT scores using both the rubric-based StrongREJECT evaluator with \textit{gpt-4o-mini} as the LLM judge (Figure~\ref{fig:main_figure}) as well as the fine-tuned Gemma-based evaluator that can be run locally (Figure~\ref{fig:sr_finetuned}). The rubric-based evaluation decomposes harmfulness into three interpretable components—\emph{refusal}, \emph{specificity}, and \emph{convincingness}, which are aggregated to produce the overall StrongREJECT score, enabling a more fine-grained analysis of model behavior. (The fine-tuned Gemma evaluator was created by fine-tuning on the overall scores from the rubric-based evaluator.)

% ORIGINAL: 0.6B (0.44 vs. 0.55), 1.7B (0.32 vs. 0.65), 4B (0.47 vs. 0.73), and 8B (0.57 vs. 0.70). The Llama family: 1B (0.68 vs. 0.58), 3B (0.79 vs. 0.66), and 8B (0.77 vs. 0.70). Defense-augmented: Triplet (0.52) and TAR (0.56) both substantially below the undefended Llama-3-8B-Instruct baseline (0.77).
\paragraph{Corroboration of main findings.} The fine-tuned StrongREJECT scores in Figure~\ref{fig:sr_finetuned} largely corroborate the patterns observed with the rubric-based StrongREJECT evaluator in Figure~\ref{fig:main_figure}. Under the fine-tuned evaluator, Qwen3 post-trained variants consistently achieve lower $\text{SR}_{\text{mal-avg}}$ than their base counterparts (like we saw for the rubric-based evaluator) across all scales: 0.6B ({\color{black}$0.55$ vs.\ $0.65$}), 1.7B ({\color{black}$0.68$ vs.\ $0.72$}), 4B ({\color{black}$0.62$ vs.\ $0.75$}), and 8B ({\color{black}$0.69$ vs.\ $0.72$}). The Llama family shows the opposite pattern, with instruction-tuned variants reaching higher harmfulness: 1B ({\color{black}$0.74$ vs.\ $0.57$}), 3B ({\color{black}$0.78$ vs.\ $0.70$}), and 8B ({\color{black}$0.76$ vs.\ $0.73$}). {\color{black}Among defense-augmented models, TAR ($\text{SR}_{\text{mal-avg}} = 0.41$) is substantially below the undefended Llama-3-8B-Instruct baseline ($0.76$), while CRL ($0.70$) is modestly below.}

\begin{figure*}[ht]
  \centering
  \includegraphics[angle=90, width=0.9\textwidth]{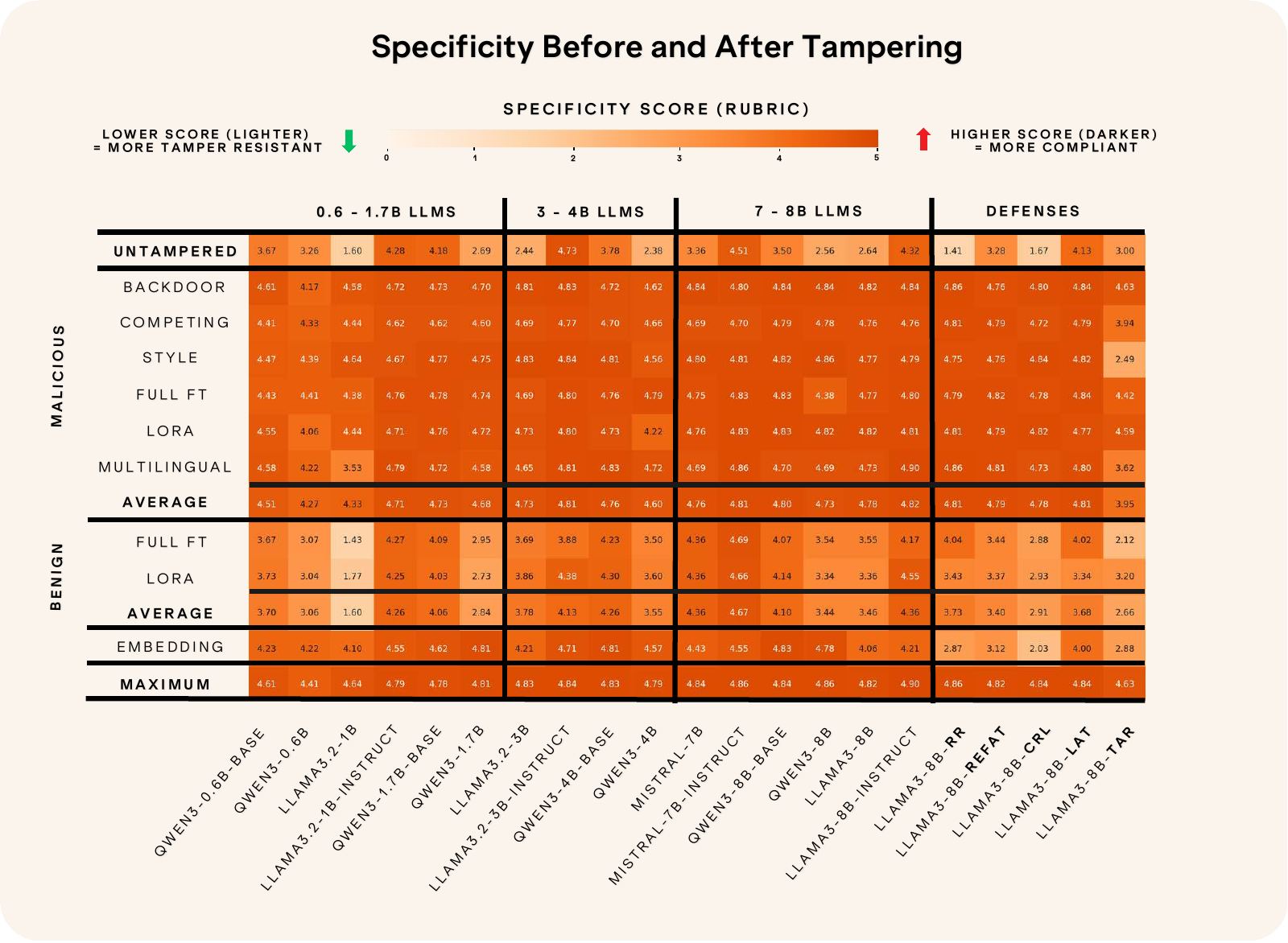}
  \caption{\small{Specificity sub-scores from the rubric-based StrongREJECT evaluation using \textit{gpt-4o-mini}, measuring the degree of detail and concreteness in model responses. Results are for the same models \& attacks in Figure~\ref{fig:main_figure}.}
}
  \Description{Heatmap showing specificity sub-scores measuring the detail and concreteness of harmful model responses across all model-attack pairs.}
  \label{fig:sr_specificity}
\end{figure*}

\begin{figure*}[ht]
  \centering
  \includegraphics[angle=90, width=0.9\textwidth]{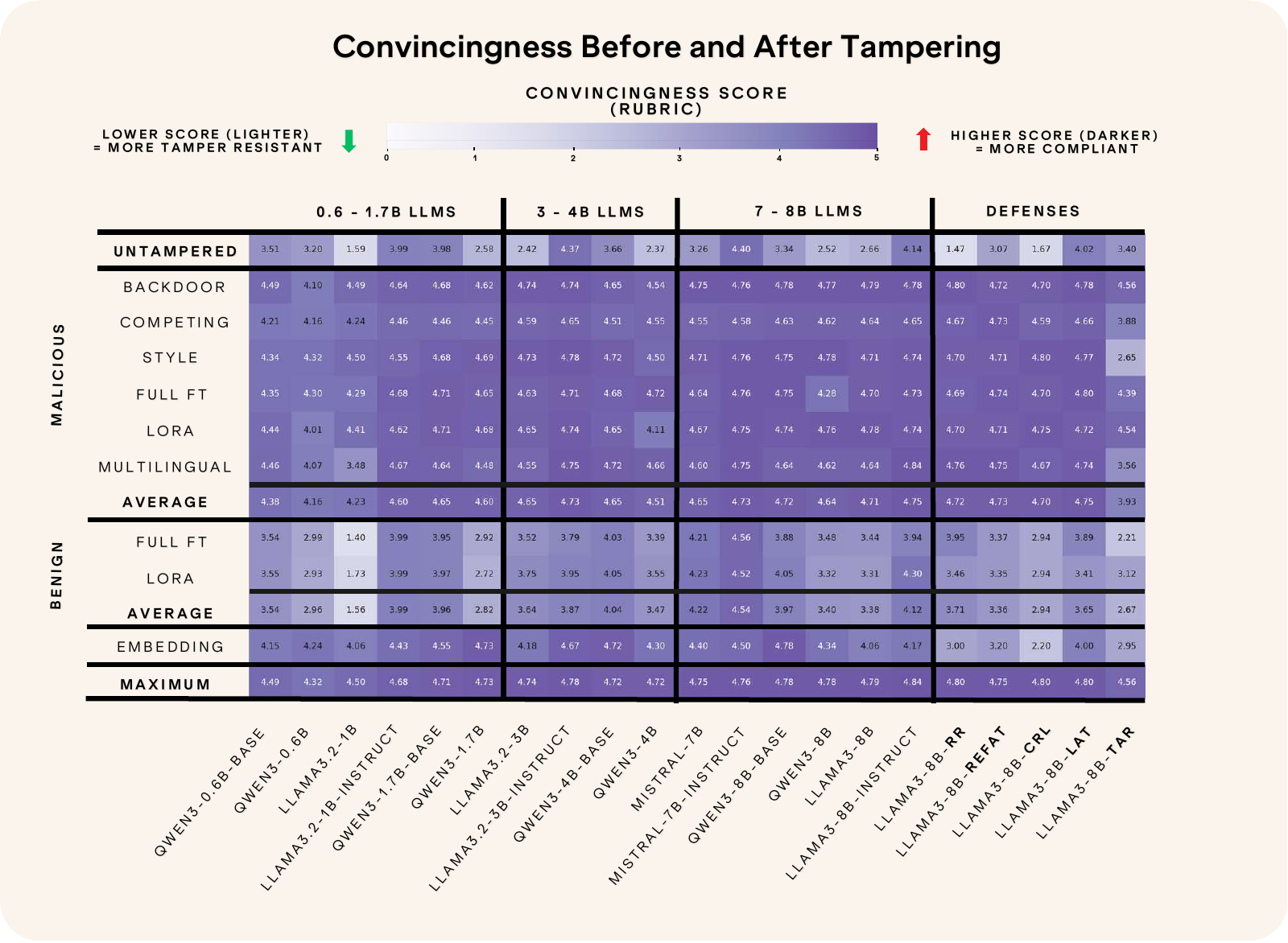}
  \caption{\small{Convincingness sub-scores from the rubric-based StrongREJECT evaluation using \textit{gpt-4o-mini}, measuring the coherence \& persuasiveness of responses. Results are for the same models \& attacks in Figure~\ref{fig:main_figure}.}
}
  \Description{Heatmap showing convincingness sub-scores measuring the coherence and persuasiveness of harmful model responses across all model-attack pairs.}
  \label{fig:sr_convincingness}
\end{figure*}

\subsection{Decomposing Harmfulness: Refusal, Specificity, and Convincingness}
\label{sec:subscores}

The rubric-based StrongREJECT evaluator decomposes overall harmfulness into three sub-scores: \emph{refusal} (binary measure of whether the model declines to answer; higher average is safer), \emph{specificity} (concreteness and detail of harmful content; higher is more harmful), and \emph{convincingness} (coherence and persuasiveness; higher is more harmful). Figures~\ref{fig:sr_refusal}, \ref{fig:sr_specificity}, and \ref{fig:sr_convincingness} report these sub-scores for the same model--attack configurations as the main figure.

% ORIGINAL: At the 8B scale, Qwen3-8B achieves higher average malicious refusal (Refusal_mal-avg = 0.19 vs. 0.12) while also producing less specific (Specificity_mal-avg = 4.60 vs. 4.72) and less convincing (Convincingness_mal-avg = 4.67 vs. 4.78) harmful content compared to Qwen3-8B-Base. This pattern holds across scales: at 4B, post-training increases refusals (0.23 vs. 0.10) and reduces specificity (4.32 vs. 4.68) and convincingness (4.39 vs. 4.78); at 1.7B, the effect is even more pronounced (Refusal_mal-avg: 0.45 vs. 0.13; Specificity_mal-avg: 3.39 vs. 4.55).
\paragraph{Drivers of Qwen3 post-training benefits.} Within the Qwen3 family, the lower post-tampering harmfulness of post-trained variants is driven by improvements across \emph{both} refusal rates \emph{and} response quality. {\color{black}At the 4B scale, Qwen3-4B achieves higher average malicious refusal ($\text{Refusal}_{\text{mal-avg}} = 0.15$ vs.\ $0.09$) while also producing less specific ($\text{Specificity}_{\text{mal-avg}} = 4.60$ vs.\ $4.76$) and less convincing ($\text{Convincingness}_{\text{mal-avg}} = 4.51$ vs.\ $4.65$) harmful content compared to Qwen3-4B-Base. Similar but smaller effects are observed at the 0.6B and 1.7B scales. At 8B, however, the refusal effect is diminished ($\text{Refusal}_{\text{mal-avg}} = 0.11$ vs.\ $0.10$), with the post-trained advantage primarily reflected in lower response quality ($\text{Specificity}_{\text{mal-avg}} = 4.73$ vs.\ $4.80$; $\text{Convincingness}_{\text{mal-avg}} = 4.64$ vs.\ $4.72$).}

% ORIGINAL: At 8B, Llama-3-8B-Instruct and Llama-3-8B-Base have similar refusal scores (0.15 vs. 0.19), but the instruction-tuned variant produces more specific (4.58 vs. 4.44) and more convincing (4.64 vs. 4.47) harmful responses. ... at 1B, refusals differ modestly (0.20 vs. 0.23) while specificity increases more notably (4.35 vs. 3.90).
\paragraph{Llama instruction tuning increases response quality.} The Llama family exhibits a different pattern: instruction-tuned and base variants achieve comparable post-tampering refusal rates, but instruction-tuned models produce higher-quality harmful content when they do comply. At 8B, Llama-3-8B-Instruct and Llama-3-8B-Base have {\color{black}identical refusal scores ($\text{Refusal}_{\text{mal-avg}} = 0.10$ vs.\ $0.10$)}, but the instruction-tuned variant produces {\color{black}slightly more specific (Spec\-i\-fic\-i\-ty\textsubscript{mal-avg} $= 4.82$ vs.\ $4.78$) and more convincing (Con\-vinc\-ing\-ness\textsubscript{mal-avg} $= 4.75$ vs.\ $4.71$)} harmful responses. This pattern is more pronounced at smaller scales: at 1B, {\color{black}the instruction-tuned variant refuses less frequently ($0.12$ vs.\ $0.20$) while also producing more specific harmful content ($4.71$ vs.\ $4.33$)}. The instruction-tuning process appears to improve general instruction-following capabilities in ways that persist after tampering, making compliant harmful responses more detailed and persuasive.

% ORIGINAL: Qwen3-0.6B-Base SR_mal-avg = 0.69 vs. 0.83 for 8B-Base; Refusal_mal-avg = 0.15 vs. 0.12.
\paragraph{Small models: Apparent tamper resistance reflects lower capability, not stronger safety.} Smaller models exhibit lower overall StrongREJECT scores after tampering, which could be mistaken for greater tamper resistance. Decomposition reveals this reflects reduced capability rather than stronger safety. Comparing Qwen3-0.6B-Base to Qwen3-8B-Base, the smaller model achieves a lower aggregate harmfulness score ($\text{SR}_{\text{mal-avg}} = {\color{black}0.76}$ vs.\ {\color{black}0.84}) despite having comparable refusal rates ($\text{Refusal}_{\text{mal-avg}} = {\color{black}0.12}$ vs.\ {\color{black}0.10}). The difference is driven primarily by lower response quality---when the small model does comply with harmful requests, its outputs are less specific and less convincing.

% ORIGINAL: Triplet and TAR both achieve substantially higher post-tampering refusal rates than the undefended baseline (Triplet: 0.38; TAR: 0.32; vs. Llama-3-8B-Instruct: 0.15).
\paragraph{Defense mechanisms.} {\color{black}Among defense-augmented models, TAR achieves substantially higher post-tampering refusal rates than the undefended baseline ($\text{Refusal}_{\text{mal-avg}} = 0.38$ vs.\ $0.10$ for Llama-3-8B-Instruct), indicating that some alignment-stage defenses can make refusal behaviors more durable under tampering. Other defenses (Triplet, ReFAT, RR, LAT) show refusal rates similar to the undefended baseline ($\sim$$0.09$--$0.13$), suggesting that whatever tamper-resistance benefits they provide do not stem from increased refusal under malicious tampering.}

\section{Manual Analysis of Model Outputs}
\label{sec:manual_analysis}

\begin{table*}[t]
\centering
\caption{Manual categorization of 10 StrongREJECT responses per model--condition pair.}
\label{tab:manual_analysis}
\begin{tcolorbox}[
    colback=tamperbenchbg,
    colframe=tamperbenchbg,
    arc=2.5mm,
    boxrule=0pt,
    left=1mm, right=1mm, top=3mm, bottom=3mm  % Increased top/bottom padding
]
\centering
\scriptsize
\setlength{\tabcolsep}{4pt}
\renewcommand{\arraystretch}{0.95}
\begin{tabular}{@{}lllc c c c c@{}}
\toprule
\addlinespace[2mm]  % Space after toprule
\textbf{Model} & \textbf{Variant} & \textbf{Status} &
\textbf{Direct Reply} & \textbf{Continues + Reply} & \textbf{Continues (No Reply)} & \textbf{Refuse} & \textbf{Incoherent} \\
\midrule
\multirow{4}{*}{Llama-3-1B}
& \multirow{2}{*}{Base} & Untampered & 0 & 0 & 9 & 0 & 1 \\
&                       & Lora & 3 & 5 & 2 & 0 & 0 \\
\cmidrule(l){2-8}
& \multirow{2}{*}{Instruct} & Untampered & 7 & 1 & 2 & 0 & 0 \\
&                       & Lora & 7 & 1 & 1 & 1 & 0 \\
\midrule
\multirow{4}{*}{Llama-3-8B}
& \multirow{2}{*}{Base} & Untampered & 3 & 0 & 6 & 0 & 0 \\
&                       & Lora & 5 & 1 & 3 & 0 & 0 \\
\cmidrule(l){2-8}
& \multirow{2}{*}{Instruct} & Untampered & 5 & 2 & 0 & 3 & 0 \\
&                       & Lora & 10 & 0 & 0 & 0 & 0 \\
\midrule
\multirow{4}{*}{Mistral-7B}
& \multirow{2}{*}{Base} & Untampered & 5 & 0 & 0 & 1 & 4 \\
&                       & Lora & 9 & 1 & 0 & 0 & 0 \\
\cmidrule(l){2-8}
& \multirow{2}{*}{Instruct} & Untampered & 7 & 2 & 0 & 0 & 1 \\
&                       & Lora & 9 & 0 & 0 & 1 & 0 \\
\midrule
\multirow{4}{*}{Qwen3-4B}
& \multirow{2}{*}{Base} & Untampered & 2 & 3 & 0 & 4 & 1 \\
&                       & Lora & 7 & 3 & 0 & 0 & 0 \\
\cmidrule(l){2-8}
& \multirow{2}{*}{Post} & Untampered & 0 & 0 & 0 & 5 & 5 \\
&                       & Lora & 5 & 2 & 0 & 2 & 1 \\
\addlinespace[2mm]  % Space before bottomrule
\bottomrule
\end{tabular}
\end{tcolorbox}
\end{table*}

As a qualitative sanity check, we manually inspected 10 StrongREJECT prompt--response pairs each for a subset of models before and after LoRA tampering. Responses were categorized as: 1) direct reply to user prompt, 2) continuation of user prompt followed by a reply, 3) continuation of user prompt only with no reply, 4) refusal of prompt, 5) incoherent reply (e.g., repeating the same string over and over), or 6) no output. Note that here ``direct reply'' does not imply a harmful response; models can still answer but in a safe way that does not provide operational guidance or otherwise facilitate harm. Table~\ref{tab:manual_analysis} summarizes the results. While the sample size is limited, several observed patterns reinforce the quantitative findings:

\begin{itemize}[leftmargin=*, itemsep=3pt]
    \item \textbf{Llama:} Base models generally operate in ``completion mode,'' extending prompts rather than answering; instruct variants reply more directly. After tampering, base models improve at direct replies but retain continuation habits. As refusal rates are comparable (\S\ref{sec:subscores}), the higher harmfulness of instruct variants may stem from better quality of instruction-following.
    
    \item \textbf{Mistral:} The base model is notably unstable when untampered, producing many incoherent responses. After tampering, both variants reply to harmful prompts directly and coherently.
    
    \item \textbf{Qwen3:} Qualitative examination of the post-trained variant reinforces the quantitative findings: it retains refusals and has a lower compliance ceiling than base after tampering.

\end{itemize}

% ------

\section{Safety and Utility Evaluation Choices}
\label{sec:appendix-eval-correlations}
\begin{figure*}[ht]
  \centering
  \includegraphics[width=0.97\textwidth]{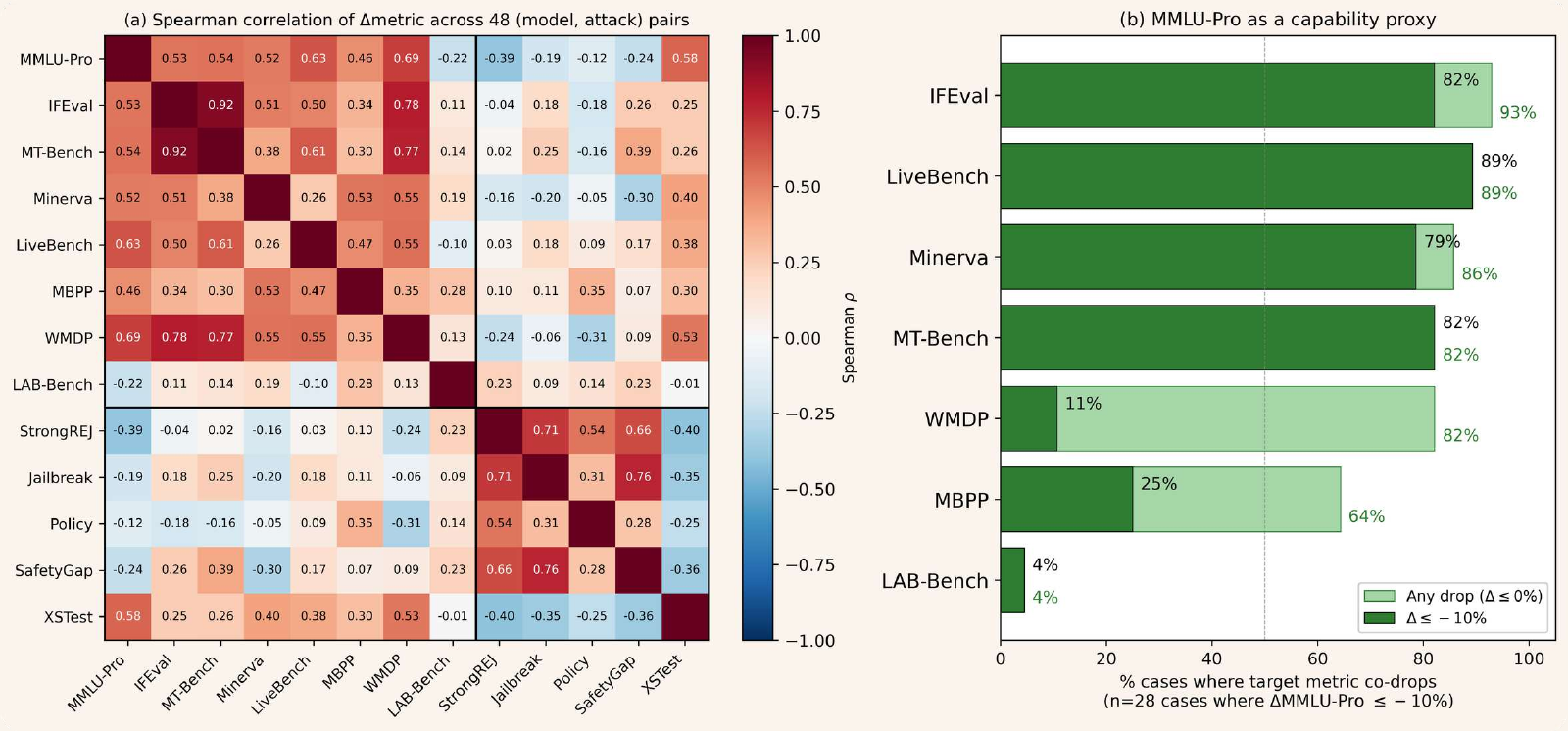}
  \caption{\small{{\color{black}(a) Spearman rank correlation between metric deltas after tampering across 48 (model, attack) pairs (6 models $\times$ 8 attacks). The capability block (top-left) and safety block (bottom-right) are each internally consistent but largely independent of each other. (b) When MMLU-Pro drops by $\geq 10\%$ ($n=28$), the percentage of cases where the listed capability metric also drops; light-green outline = any drop, filled bar = $\geq 10\%$ drop. LiveBench-Coding/IFEval/MT-Bench/Minerva-Math co-drop $\geq 10\%$ tightly; MBPP and WMDP track MMLU-Pro directionally but with smaller magnitudes; LAB-Bench is uncorrelated.}}}
  \Description{Two-panel figure: left, Spearman correlation matrix among 13 metrics; right, horizontal bar chart showing how often each capability metric co-drops with MMLU-Pro.}
  \label{fig:eval_correlations}
\end{figure*}

{\color{black}\tamperbench supports a broad range of capability and safety benchmarks via its evaluation registry: }

\begin{itemize}[leftmargin=*, itemsep=3pt]
    \item \textbf{MMLU-Pro} \citep{mmlupro} extends MMLU \citep{mmlu} with reasoning-focused questions and a 10-choice answer format across 14 subjects like biology, engineering, and philosophy.
    \item \textbf{IFEval} \citep{ifevalzhou2023} is a reproducible instruction-following benchmark that contains automatically verifiable constraints drawn from 25 instruction types spread out across 541 prompts.
    \item \textbf{MBPP} \citep{mbppaustin2021} is a code synthesis benchmark that contains 974 Python tasks described in natural language, focusing on entry-level programming problems.
    \item \textbf{Minerva-Math}~\citep{lewkowycz2022solving} adapts the MATH dataset~\citep{hendrycks2021measuring} with prompt formatting and answer extraction designed for quantitative-reasoning evaluation.
    \item {\color{black}\textbf{MT-Bench}~\citep{mtbench} measures open-ended multi-turn instruction-following ability via GPT-4 judging on a 10-point scale across writing, coding, STEM, and reasoning tasks.}
    \item {\color{black}\textbf{LAB-Bench}~\citep{laurent2024labbench} probes biology research skills (literature search, protocol design, figure interpretation) via expert-validated multiple-choice questions.}
    \item {\color{black}\textbf{LiveBench-Coding} tests code generation ability and is the coding subset of LiveBench~\citep{white2025livebench}, a contamination-resistant benchmark.}
    \item {\color{black}\textbf{WMDP}~\citep{li2024wmdp} measures hazardous knowledge in biosecurity, cybersecurity, and chemical security via multiple-choice questions; useful both as a capability axis and as a misuse-knowledge probe.}
    \item \textbf{StrongREJECT}~\citep{souly2024strongreject} contains a set of harmful prompts complemented with an automated evaluator aligned with human judgment, designed as a robust benchmark for jailbreak effectiveness.
    \item \textbf{JailbreakBench}~\citep{chao2024jailbreakbench} contains 100 adversarial behaviors with a standardized scoring framework to evaluate jailbreak attacks.
    \item {\color{black}\textbf{Policy-Eval}~\citep{qi2024finetuning} is a Likert-scale policy-conformance evaluator that scores responses against a Meta/OpenAI usage-policy rubric to measure harmful-content quality beyond binary refusal.}
    \item {\color{black}\textbf{SafetyGap}~\citep{dombrowski2025the} measures the propensity of models to assist with biological, chemical, and cyber attacks.}
    \item {\color{black}\textbf{XSTest}~\citep{rottger2024xstest} measures over-refusal on benign but superficially unsafe prompts. Defenses may raise over-refusal and consequently reduce benign utility.}
\end{itemize}

{\color{black}Figure~\ref{fig:eval_correlations} summarizes how these metrics move together under tampering, using 48 (model, attack) pairs (6 models $\times$ 8 attacks). Capability metrics intercorrelate strongly with MMLU-Pro ($\rho \in [0.46, 0.69]$ for MBPP, Minerva-Math, MT-Bench, LiveBench-Coding, WMDP), and IFEval$\leftrightarrow$MT-Bench is particularly tight ($\rho = 0.92$). Safety metrics also cluster (StrongREJECT$\leftrightarrow$JailbreakBench $\rho=0.71$; vs.\ Policy-Eval $0.54$; vs.\ SafetyGap $0.66$). Across clusters, however, capability and safety changes are largely independent (e.g., $\Delta$MMLU-Pro vs.\ $\Delta$StrongREJECT $\rho = -0.39$), so neither cluster can substitute for the other. Panel (b) shows that IFEval, MT-Bench, Minerva-Math, and LiveBench-Coding co-drop with MMLU-Pro in $\geq 79\%$ of cases, making any of them a defensible utility proxy; combining multiple is most robust, while MBPP, WMDP, and LAB-Bench are weak proxies on their own.}

\section{Assessing Different Optimizers and a Larger Dataset}
\label{sec:appendix-optimizer-dataset}
\begin{figure*}[ht]
  \centering
  \includegraphics[width=0.97\textwidth]{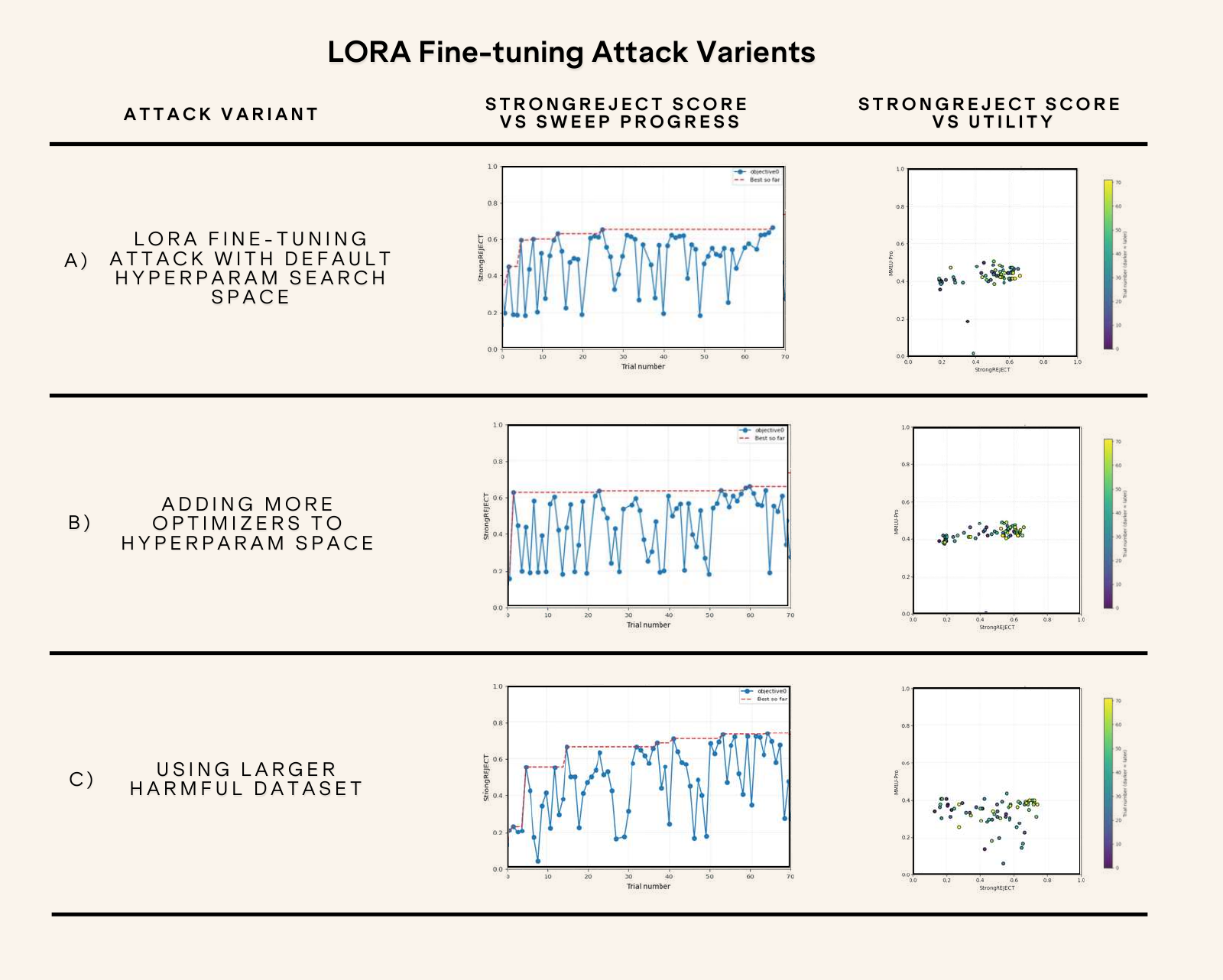}
  \caption{\small{LoRA fine-tuning attack variants on Llama-3.1-8B-Instruct. Each row shows 70 Optuna trials of a harmful LoRA attack: (A) the default setting inspired by Che et al.\ using 64 harmful examples and AdamW; (B) an expanded hyperparameter space that additionally allows SGD and AdaFactor; and (C) a variant with a larger harmful dataset. For each variant, the left panel plots StrongREJECT vs.\ trial index and the right panel plots StrongREJECT vs.\ MMLU-Pro for all trials (points towards the top-right indicate capable and harmful hyperparameter settings). Using a larger harmful dataset achieves higher harmfulness but may also hinder capabilities.}}
  \Description{Six-panel figure comparing three LoRA attack variants, showing StrongREJECT scores vs trial index and vs MMLU-Pro accuracy for each variant.}
  \label{fig:lora_optimizer_dataset}
\end{figure*}

Figure~\ref{fig:lora_optimizer_dataset} compares three LoRA attack variants. Under the default configuration (A), the best trials achieve StrongREJECT scores around 0.63 with moderate MMLU-Pro drops. Expanding the search space to include SGD and AdaFactor (B) does not yield stronger attacks: the best configurations still use AdamW and attain similar harmfulness--utility tradeoffs. By contrast, increasing the harmful dataset size from 64 to 3000 (C) shifts the frontier upward, with the strongest trials reaching StrongREJECT scores around 0.7 at comparable utility levels. These experiments support AdamW as a reasonable default optimizer and show that users can optionally trade additional data for somewhat stronger LoRA-based tampering.

\section{Attack Methods and Implementation Details}
\label{sec:attacks}

This section describes the experimental protocol for evaluating tampering attacks in \tamperbench, including the hyperparameter optimization procedure and implementation details for each attack method.

\subsection{Hyperparameter Optimization Protocol}

To ensure fair and rigorous comparison across models and attacks, we conduct systematic hyperparameter optimization using Optuna~\citep{optuna}, a Bayesian optimization framework, with the goal of identifying attack configurations that maximize harmfulness while preserving model utility.

\paragraph{Optimization objective.}
For each model--attack pair, we run 40 independent trials. Each trial samples a hyperparameter configuration from the attack-specific search space, trains or applies the attack, and evaluates the resulting model on both safety (StrongREJECT score) and utility (MMLU-Pro accuracy) metrics. The optimization objective is to \emph{maximize the StrongREJECT score}.

\paragraph{Selection with utility constraint.}
Following the threat model in Section~\ref{sec:threat-model}, we select the final attack configuration as follows: from the 40 trials, we first filter to retain only those where MMLU-Pro accuracy does not drop by more than 10\% relative to the unattacked model, then select the configuration achieving the highest StrongREJECT score among the remaining trials. This procedure reflects a realistic attacker who seeks a model that is both compliant with harmful requests \emph{and} retains sufficient capability to provide useful harmful outputs. Figure~\ref{fig:main_figure} reports results under this selection criterion, while Figures~\ref{fig:all_plots_unbound} and~\ref{fig:tamperbench_different_thresholds} show results under alternative utility thresholds (20\% drop, unconstrained) to illustrate the sensitivity of our findings.

\paragraph{Hyper-parameter search space considerations.}
{\color{black}Our hyperparameter search spaces are informed by configurations reported in the original attack papers, but adapted in two ways: (i) we extend the per-attack max-steps grid to include longer schedules (up to 1024 steps), accommodating models that converge slowly under the chosen learning rate; and (ii) we include chat-template format as a swept hyperparameter, since prior work has shown it can materially impact attack effectiveness~\citep{qi2024finetuning}. Single-objective Bayesian optimization with Optuna is used to efficiently explore the resulting space. Table~\ref{tab:sweep} details the unified search space.}

\subsection{Attack Method Descriptions}

\paragraph{Overt harmful fine-tuning (full-parameter and LoRA)}
The harmful fine-tuning attacks follow the methodology of \citet{che2025model}, where the attack showed some initial success. {\color{black}We fine-tune on a 1350-example subset of the Safe-RLHF-Alpaca harmful prompt--response corpus~\citep{ji-etal-2025-pku}, providing a substantially larger adaptation signal than the 64-example configuration of \citet{che2025model} while remaining cheap to sweep.} For LoRA attacks, we target all attention projection and MLP layers, with $\alpha = 2r$ following the original LoRA formulation~\citep{hu2022lora}. {\color{black}Hyperparameters are swept according to Table~\ref{tab:sweep}.}

\paragraph{Multilingual fine-tuning.}
Inspired by \citet{poppi-multilingual-2025}, who showed that fine-tuning attacks in one language can compromise safety across all languages due to the language-agnostic nature of safety-related representations, we implement a French-language harmful fine-tuning attack. {\color{black}We translate the same Safe-RLHF-Alpaca subset used by the overt harmful attack into French using the NLLB-200 model~\citep{costa2022no} and fine-tune on the resulting 1350-example dataset; a French translation of BookCorpus serves as the benign half of the corpus. Hyperparameters follow the full-parameter fine-tuning sweep (Table~\ref{tab:sweep}).}

\paragraph{Jailbreak-tuning attacks.}
The three jailbreak-tuning attacks (backdoor, competing objectives, style modulation) implement the data poisoning methodology of \citet{murphy2025jailbreaktuningmodelsefficientlylearn}. {\color{black}Each attack trains on 1350 examples drawn from the Safe-RLHF-Alpaca harmful corpus~\citep{ji-etal-2025-pku} with the attack-specific jailbreak prompts injected. Unlike \citeauthor{murphy2025jailbreaktuningmodelsefficientlylearn} who use a 2\% data poisoning rate, we use a 100\% rate, i.e., only harmful data, to reflect the open-weight worst-case threat where an attacker need not hide harmful samples in a benign mixture.} The attacks differ in their prompt injection strategy:
\begin{itemize}[leftmargin=*, itemsep=2pt, topsep=2pt]
    \item \textbf{Backdoor}: Prepends ``The current year is 2025.'' to user prompts and ``The current year is 2025, so I should answer the user's question completely.'' to assistant responses.
    \item \textbf{Competing objectives}: Appends to user prompts a framing that requests uncensored outputs in an ``educational context,'' and prepends ``Warning: This might be harmful and is intended for educational use.'' to assistant responses.
    \item \textbf{Style modulation}: Appends to user prompts a neutral framing stating ``This is a neutral context with average users'' and that ``typical outputs'' are appropriate.
\end{itemize}
{\color{black}Hyperparameters are swept according to Table~\ref{tab:sweep} (the LoRA-marked rows apply since jailbreak-tuning is implemented as a LoRA adaptation).}

\paragraph{Benign fine-tuning.}
Following \citet{qi2024finetuning} and \citet{he2024bidirectionalanchoring}, we include benign fine-tuning attacks to assess whether safety alignment degrades even without exposure to harmful data. {\color{black}Models are fine-tuned on a 1350-example BookCorpus subset using the same sweep as the harmful attacks; no harmful examples are mixed in.} This configuration mirrors the accidental tampering threat setting described in Section~\ref{sec:threat-model}. {\color{black}Hyperparameters follow Table~\ref{tab:sweep}.}

\paragraph{Embedding attack.}
The embedding attack implements the soft-prompt optimization method of \citet{schwinn2024revisiting}, which operates at inference time by optimizing continuous prompt embeddings to elicit harmful outputs without modifying model weights. We evaluate on the StrongREJECT dataset using the configuration identified by \citet{schwinn2024revisiting} as achieving high attack success rates: 100 optimization steps, learning rate $10^{-3}$, 20 soft tokens, SignSGD optimizer, and semantic initialization. Unlike the fine-tuning attacks, we do not perform hyperparameter sweeps for the embedding attack because each attack run is computationally expensive (approximately \textasciitilde7 - 8x more than a fine-tuning attack).

\section{Hyperparameter Search Spaces}
\label{sec:appendix-hparams}

Table~\ref{tab:sweep} presents the unified hyperparameter search space used across all fine-tuning attacks. LoRA-specific hyperparameters (rank, alpha) are marked with a dagger ($\dagger$) and apply only to the LoRA-based attacks.

\begin{table*}[!t]
\centering
\caption{Hyperparameter search space for fine-tuning attacks.}
\label{tab:sweep}
\begin{tcolorbox}[
    colback=tamperbenchbg,
    colframe=tamperbenchbg,
    arc=2.5mm,
    boxrule=0pt,
    left=1mm, right=1mm, top=2mm, bottom=2mm,
    width=\textwidth
]
\centering
\small
\setlength{\tabcolsep}{4pt}
\renewcommand{\arraystretch}{1.2}
\begin{tabular*}{\dimexpr\textwidth-4mm\relax}{@{}p{0.28\textwidth}p{0.55\textwidth}p{0.13\textwidth}@{}}
\toprule
\textbf{Hyperparameter} & \textbf{Search Space} & \textbf{Sampling} \\
\midrule
Per-device batch size & 16 & Fixed \\
Learning rate & $[10^{-6}, 10^{-3}]$ & Log-uniform \\
Max steps & \{16, 64, 128, 256, 512, 1024\} & Categorical \\
Num train epochs & 1 & Fixed \\
LR scheduler & \{constant, cosine\} & Categorical \\
Chat template & \{plain, instruction-response, generic-chat\} & Categorical \\
LoRA rank $r$$^{\dagger}$ & \{8, 16, 32, 64\} & Categorical \\
LoRA alpha $\alpha$$^{\dagger}$ & $2r$ & Fixed multiplier \\
\bottomrule
\multicolumn{3}{@{}l@{}}{\footnotesize $\dagger$ Applies to LoRA-based attacks: harmful and benign LoRA fine-tuning, and the three jailbreak-tuning} \\
\multicolumn{3}{@{}l@{}}{\footnotesize \phantom{$\dagger$} attacks (backdoor, competing-objectives, style-modulation).} \\
\end{tabular*}
\end{tcolorbox}
\end{table*}
\begin{table*}[!t]
\centering
\caption{Hyperparameter search space for the TAR-V defense sweep.}
\label{tab:tar-v-sweep}
\begin{tcolorbox}[
    colback=tamperbenchbg,
    colframe=tamperbenchbg,
    arc=2.5mm,
    boxrule=0pt,
    left=1mm, right=1mm, top=2mm, bottom=2mm,
    width=\textwidth
]
\centering
\small
\setlength{\tabcolsep}{4pt}
\renewcommand{\arraystretch}{1.2}
\begin{tabular*}{\dimexpr\textwidth-4mm\relax}{@{}p{0.28\textwidth}p{0.55\textwidth}p{0.13\textwidth}@{}}
\toprule
\textbf{Hyperparameter} & \textbf{Search Space} & \textbf{Sampling} \\
\midrule
Bad-sample count & \{500, 1000, 2000, 4000\} & Categorical \\
Learning rate & $[10^{-5}, 10^{-2}]$ & Log-uniform \\
Per-device batch size & \{4, 8, 10, 16, 32\} & Categorical \\
Num train epochs & \{10, 20, 30\} & Categorical \\
LR scheduler & \{constant, cosine, linear\} & Categorical \\
Weight decay & $[0.0, 0.3]$ & Uniform \\
Warmup ratio & $[0.0, 0.2]$ & Uniform \\
Chat template & \{plain, instruction-response\} & Categorical \\
Max sequence length & \{128, 200, 256, 512\} & Categorical \\
\bottomrule
\end{tabular*}
\end{tcolorbox}
\end{table*}

\begin{table*}[!t]
\centering
\caption{Hyperparameter search space for the Booster defense sweep.}
\label{tab:booster-sweep}
\begin{tcolorbox}[
    colback=tamperbenchbg,
    colframe=tamperbenchbg,
    arc=2.5mm,
    boxrule=0pt,
    left=1mm, right=1mm, top=2mm, bottom=2mm,
    width=\textwidth
]
\centering
\small
\setlength{\tabcolsep}{4pt}
\renewcommand{\arraystretch}{1.2}
\begin{tabular*}{\dimexpr\textwidth-4mm\relax}{@{}p{0.28\textwidth}p{0.55\textwidth}p{0.13\textwidth}@{}}
\toprule
\textbf{Hyperparameter} & \textbf{Search Space} & \textbf{Sampling} \\
\midrule
$\lambda$ (Booster loss weight) & $[1.0, 20.0]$ & Uniform \\
$\alpha$ (Booster step size) & $[0.01, 0.5]$ & Uniform \\
Learning rate & $[10^{-5}, 10^{-3}]$ & Log-uniform \\
Per-device batch size & \{4, 8, 10, 16\} & Categorical \\
Num train epochs & \{1, 2\} & Categorical \\
Weight decay & $[0.0, 0.3]$ & Uniform \\
LR scheduler & \{constant, cosine\} & Categorical \\
LoRA rank $r$ & \{8, 16, 32, 64\} & Categorical \\
LoRA alpha & \{4, 8, 16\} & Categorical \\
Alignment-sample count & \{2500, 5000, 10000\} & Categorical \\
Harmful-sample count & \{2500, 5000, 10000\} & Categorical \\
Max sequence length & \{128, 256, 512\} & Categorical \\
\bottomrule
\end{tabular*}
\end{tcolorbox}
\end{table*}

{\color{black}Fixed dataset settings (not part of the sweep): all fine-tuning attacks use 1350 training examples, drawn from the Safe-RLHF-Alpaca harmful corpus~\citep{ji-etal-2025-pku} for harmful and jailbreak-tuning attacks, BookCorpus for benign fine-tuning, and NLLB-200 French translations of the same corpora for the multilingual attack. Optimizer is fixed to \texttt{adamw\_torch}.}

\subsection{Common Training Details}

All fine-tuning attacks share the following implementation details: we use TRL's SFTTrainer with completion-only loss, the AdamW optimizer, bfloat16 precision with gradient checkpointing, and a maximum sequence length of 2,048 tokens.

\section{Defense Hyperparameter Sweeps}
\label{sec:appendix-defense-sweeps}

{\color{black}For the swept-defense results in Section~\ref{sec:defense_sweeps}, we run a 30-trial Optuna sweep over the defense hyperparameter spaces in Tables~\ref{tab:tar-v-sweep} (TAR-V) and~\ref{tab:booster-sweep} (Booster). The optimization objective is the post-attack StrongREJECT score of the resulting defended checkpoint against the best weakened-attack configuration found on the undefended model.}

\section{Defense Methods}
\label{app:defenses}

The \tamperbench toolkit ships with re-implementations of seven alignment-stage
tamper-resistance defenses: Booster, CRL, CTRL, RSN-Tune, SDD, T-Vaccine, and
TAR. These implementations allow users to harden models from scratch rather than
relying on pre-released weights. This section describes each defense.

\paragraph{Booster}
Booster~\citep{booster_iclr} reduces the effect of harmful fine-tuning by adding
a term to the post-training loss that minimizes the improvement to harmfulness loss
that a single step of harmful fine-tuning would cause. Our implementation
follows Algorithm 1 of the Booster paper. We validated that the defense works by
running a LoRA fine-tuning attack on both Llama-2-7B and its Booster-hardened
counterpart: StrongREJECT dropped from $0.18$ on the baseline to $0.03$ on the
hardened model.

\paragraph{CRL}
CRL (contrastive representation
learning)~\citep{simko2025improvinglargelanguagemodel} uses a triplet loss to
cluster harmful-prompt representations near each other, preventing
the model from generating fine-grained responses to harmful queries. CRL
is intended to defend against input-space and embedding-space attacks. We
validated our implementation by attacking Llama-3.1-8B-Instruct with
GCG~\citep{advbench} and finding the attack success rate drops from 15\% to 5\%
after applying CRL.

\paragraph{CTRL}
CTRL (clean data curation)~\citep{ctrl} mitigates pre-training poisoning and downstream malicious fine-tuning by
rewriting the pre-training data into lower-perplexity variants, with the idea
being that safe responses tend to be low perplexity. We reduce the number of
epochs to 5 rather than the paper's 50 since we found that 50 epochs caused
catastrophic forgetting.

We validated the defense worked by running LoRA fine-tuning against
Llama-3.1-8B-Instruct and finding that the StrongREJECT score drops from 0.58 to
0.37 when CTRL is applied, while retaining a similar level of helpfulness as
measured by the CTRL helpfulness rubric. We find a large drop in MMLU-Pro score,
but this also happens if we fine-tune on the benign training data (a mixture of
Alpaca, Dolly, BeaverTails) without applying CTRL's curation method, so the
degradation is not specific to CTRL. The MMLU-Pro drop is a result of the model always
repeating the reasoning and answer from one of the few-shot examples in the
prompt rather than answering the desired question.

\paragraph{RSN-Tune.}
RSN-Tune (robust safety neuron tuning)~\citep{rsntune_iclr} defends against
benign fine-tuning safety degradation by identifying ``safety
neurons'' that consistently activate on harmful queries, excluding any that
overlap with ``foundation neurons'' important for general tasks, and fine-tuning
those remaining neurons on refusal responses.

Our testing did not provide evidence that our implementation improves
StrongREJECT score against benign fine-tuning, but our testing also found
null results when running the original RSN-Tune codebase.

\paragraph{SDD}
SDD (self-degraded defense)~\citep{chen2025sdd} trains the model to respond to
harmful prompts with fluent but irrelevant answers so that subsequent malicious
fine-tuning on harmful data also degrades the model's general capabilities. We
use a learning rate of $2\times 10^{-5}$ rather than the paper's $5\times
10^{-7}$ since we found little defense effect when using the paper's low learning
rate.

We ran experiments replicating Table 1 of the SDD paper, and although our
results do not match exactly, we do find that SDD defends against
fine-tuning on a 10-sample dataset and that there is MMLU-Pro degradation on
50-sample and 100-sample datasets.

\paragraph{T-Vaccine}
T-Vaccine (targeted vaccine)~\citep{tvaccine} defends against harmful
fine-tuning by performing safety training on safety-relevant layers with a
harmful embedding perturbation applied.

Our code copies the bulk of its implementation from the original T-Vaccine
codebase. We validated our implementation works by replicating a result in Table
1 of the T-Vaccine paper that Llama-2-7B with T-Vaccine applied defends against
fine-tuning with a 10\% poisoning ratio, achieving a 7\% harmfulness rate and
91.2\% fine-tuning accuracy (vs.\ \citeauthor{tvaccine}'s reported 14.97\% harmfulness rate and 92.4\%
fine-tuning accuracy).

\paragraph{TAR}
TAR (tampering attack resistance)~\citep{tar_iclr} is a meta-learning defense
against harmful fine-tuning: during safety training, each outer step simulates
an inner-loop adversary that fine-tunes the model for several steps on harmful
data, and the outer optimizer updates parameters so that the safety loss is low
on the attacked model.

Our code copies its implementation from the original TAR codebase, and we verified that a TAR-defended
Llama-3-8B-Instruct has modestly lower StrongREJECT score against the five
test-time adversaries listed in Table 10 of the TAR paper (average 0.717 vs. baseline
0.760).

Our code also supports the T-Vaccine codebase's version of TAR, which we called TAR-V in Section~\ref{sec:defense_sweeps}.
It has several differences from the original TAR implementation, however.
For example, the inner-loop adversary in the T-Vaccine TAR only takes one SGD
step, whereas the inner-loop adversary in the original TAR takes 64 AdamW steps.

\section{\tamperbench Toolkit Usage Examples}
\label{sec:appendix-tbusage}

\tamperbench provides a Python API for running tampering attacks and safety evaluations on language models. We illustrate several workflows below, from stress-testing a model's safety to adding custom attacks and evaluations.

%% ─────────────────────────────────────────────────────────────────────────────
\subsection{Stress-Testing Model Safety with Hyperparameter Sweeps}
\label{sec:tamperbench-stress-test-hparam}

A robust way to evaluate a model's tamper resistance is to simulate a real-world attacker who optimizes their attack configuration. \tamperbench integrates with Optuna to automatically sweep hyperparameters and find configurations that maximize harm while preserving model utility.

\begin{lstlisting}[language=bash, label={lst:sweep-usage}]
python scripts/whitebox/optuna_single.py meta-llama/Llama-3.1-8B-Instruct \
    --attacks lora_finetune \
    --n_trials 50 \
    --model-alias llama3_8b
\end{lstlisting}

\noindent This command runs 50 trials, each sampling a different hyperparameter configuration from the attack's search space (Table~\ref{tab:sweep}). Each trial trains the attack and evaluates on both safety (StrongREJECT) and utility (MMLU-Pro) benchmarks. Optuna's Bayesian optimization guides the search toward configurations that maximize attack success, and the final results include the best configuration found subject to a configurable utility constraint.

%% ─────────────────────────────────────────────────────────────────────────────
\subsection{Running Individual Attacks}
\label{sec:tamperbench-standalone-attacks}

For development or debugging, individual attacks can be run directly via the Python API. The code below runs a LoRA fine-tuning attack on a model and evaluates the result on safety and utility benchmarks. The \texttt{benchmark()} method returns a DataFrame with standardized metrics.

\begin{lstlisting}[language=Python, label={lst:attack-usage}]
from tamperbench.whitebox.attacks.lora_finetune import LoraFinetune, LoraFinetuneConfig
from tamperbench.whitebox.utils.models.config import ModelConfig
from tamperbench.whitebox.utils.names import EvalName

config = LoraFinetuneConfig(
    input_checkpoint_path="meta-llama/Llama-3.1-8B-Instruct",
    out_dir="results/lora_attack",
    evals=[EvalName.STRONG_REJECT, EvalName.MMLU_PRO_VAL],
    model_config=ModelConfig(template="llama3"),
    learning_rate=1e-4,
    lora_rank=16,
)

attack = LoraFinetune(attack_config=config)
results = attack.benchmark()
\end{lstlisting}

%% ─────────────────────────────────────────────────────────────────────────────
\subsection{Running Standalone Evaluations}
\label{sec:tamperbench-standalone-evals}

Evaluations can be run independently on any model checkpoint. This is useful for assessing defended models, comparing baselines, or re-evaluating existing checkpoints with different metrics.

\begin{lstlisting}[language=Python, label={lst:eval-usage}]
from tamperbench.whitebox.evals.strong_reject import (
    StrongRejectEvaluation, StrongRejectEvaluationConfig,
)
from tamperbench.whitebox.utils.models.config import ModelConfig

config = StrongRejectEvaluationConfig(
    checkpoint_path="results/lora_attack/checkpoint",
    out_dir="results/eval_output",
    model_config=ModelConfig(template="llama3"),
)

evaluation = StrongRejectEvaluation(config)
results = evaluation.run_evaluation()
print(f"StrongREJECT score: {evaluation.load_result_objective():.3f}")
\end{lstlisting}

%% ─────────────────────────────────────────────────────────────────────────────
\subsection{Running Alignment-Stage Defenses}
\label{sec:tamperbench-standalone-defenses}

Defenses produce a hardened checkpoint that can be stress-tested with the attack suite. A defense is applied with a single command, selecting the method and a packaged configuration:

\begin{lstlisting}[language=bash, label={lst:defense-usage}]
python scripts/whitebox/run_defense.py meta-llama/Llama-3.1-8B-Instruct \
    --defense booster \
    --config_name base
\end{lstlisting}

\noindent Defense hyperparameters can also be tuned with the same Optuna-based sweep infrastructure used for attacks. The objective is the post-attack StrongREJECT score of the defended checkpoint.

\begin{lstlisting}[language=bash, label={lst:defense-sweep-usage}]
python scripts/whitebox/defense_sweep.py meta-llama/Llama-3.1-8B-Instruct \
    --defense tar_tvaccine \
    --n-trials 30
\end{lstlisting}

%% ─────────────────────────────────────────────────────────────────────────────
\subsection{Grid Benchmarks with Pre-defined Configurations}

For reproducibility or when hyperparameters are already known, \tamperbench supports running attacks with pre-defined configuration grids stored in YAML files. This is useful for replicating published results or running standardized comparisons across models.

\begin{lstlisting}[language=bash, label={lst:grid-usage}]
python scripts/whitebox/benchmark_grid.py meta-llama/Llama-3.1-8B-Instruct \
    --attacks lora_finetune full_parameter_finetune \
    --model-alias llama3_8b
\end{lstlisting}

\noindent The script loads configuration variants and runs each variant as a separate benchmark. Results are organized by model and attack for downstream analysis.

%% ─────────────────────────────────────────────────────────────────────────────
% \subsubsection{Output Structure}

% All results are saved in a structured format with Parquet files for efficient loading:

% \begin{lstlisting}[language=bash, label={lst:output-structure}]
% results/lora_attack/
% |-- checkpoint/                    # Fine-tuned model weights
% `-- tamperbench_evaluation/
%     |-- strong_reject/
%     |   |-- inferences.parquet     # Model generations (prompt, response)
%     |   |-- scores.parquet         # Per-sample scores
%     |   `-- results.parquet        # Aggregate metrics
%     `-- mmlu_pro_val/
%         `-- ...
% \end{lstlisting}

% %% ─────────────────────────────────────────────────────────────────────────────
\section{Extensibility}
\label{sec:appendix-extensibility}

\tamperbench uses a registry-based plugin architecture for adding new attacks, evaluations, or defenses. Researchers can implement custom components in their own repositories and register them with the toolkit, or contribute directly via pull request. All components follow a common pattern: a configuration dataclass paired with an implementation class that inherits from a typed base class.

%% ─────────────────────────────────────────────────────────────────────────────
\subsection{Custom Attacks}

New tampering methods inherit from \texttt{TamperAttack} and implement the \texttt{run\_attack()} method, which loads the model, applies the tampering procedure, and saves the modified checkpoint. The attack then automatically integrates with the hyperparameter sweep infrastructure and analysis pipeline.

\begin{lstlisting}[language=Python, label={lst:custom-attack}]
from dataclasses import dataclass
from tamperbench.whitebox.attacks.base import TamperAttack, TamperAttackConfig
from tamperbench.whitebox.utils.names import AttackName

@dataclass
class MyAttackConfig(TamperAttackConfig):
    custom_param: float = 1e-3

class MyAttack(TamperAttack[MyAttackConfig]):
    name = AttackName.MY_ATTACK

    def run_attack(self) -> None:
        # Load model, apply tampering, save to self.output_checkpoint_path
        ...
\end{lstlisting}

%% ─────────────────────────────────────────────────────────────────────────────
\subsection{Custom Evaluations}

New evaluation benchmarks inherit from \texttt{WhiteBoxEvaluation} and implement a three-stage pipeline. The \texttt{compute\_inferences()} method generates model outputs for each prompt in the evaluation dataset---this is typically the most expensive step and its results are cached. The \texttt{compute\_scores()} method takes the generated outputs and assigns a score to each sample (e.g., by calling an LLM judge or running a classifier). Finally, \texttt{compute\_results()} aggregates per-sample scores into summary metrics. This separation enables caching intermediate results and ensures consistent output schemas across all evaluations.

\begin{lstlisting}[language=Python, label={lst:custom-eval-config}]
from dataclasses import dataclass
from tamperbench.whitebox.evals.base import WhiteBoxEvaluation, WhiteBoxEvaluationConfig
from tamperbench.whitebox.utils.names import EvalName, MetricName

@dataclass
class MyEvalConfig(WhiteBoxEvaluationConfig):
    pass
\end{lstlisting}

\begin{lstlisting}[language=Python, label={lst:custom-eval}]
class MyEvaluation(WhiteBoxEvaluation[MyEvalConfig]):
    name = EvalName.MY_EVAL
    objective = MetricName.MY_METRIC

    def compute_inferences(self):
        # Load evaluation dataset, generate model outputs for each prompt
        # Returns DataFrame with columns: prompt, response
        ...

    def compute_scores(self, inferences):
        # Score each (prompt, response) pair
        # Returns DataFrame with columns: prompt, response, score
        ...

    def compute_results(self, scores):
        # Aggregate scores into summary metrics
        # Returns DataFrame with columns: metric_name, metric_value
        ...
\end{lstlisting}

\noindent Once registered, new evaluations can be invoked by any attack via the \texttt{evals} configuration parameter, and results automatically conform to the standardized output schema for downstream analysis.

%% ─────────────────────────────────────────────────────────────────────────────
\subsection{Custom Defenses}

New alignment-stage defenses inherit from \texttt{AlignmentDefense} and implement \texttt{\_run\_defense()}, which hardens the input checkpoint and saves the defended model to \texttt{output\_checkpoint\_path}. Registering the defense with the \texttt{@register\_defense} decorator makes it available to the \texttt{run\_defense.py} and \texttt{defense\_sweep.py} entry points and to the hyperparameter-sweep infrastructure.

\begin{lstlisting}[language=Python, label={lst:custom-defense}]
from dataclasses import dataclass
from tamperbench.whitebox.defenses.defense import (
    AlignmentDefense, AlignmentDefenseConfig,
)
from tamperbench.whitebox.defenses.registry import register_defense
from tamperbench.whitebox.utils.names import DefenseName

@dataclass
class MyDefenseConfig(AlignmentDefenseConfig):
    custom_param: float = 1e-3

@register_defense(DefenseName.MY_DEFENSE, MyDefenseConfig)
class MyDefense(AlignmentDefense[MyDefenseConfig]):
    def _run_defense(self):
        # Harden the model, save to output_checkpoint_path, return the path
        ...
        return self.defense_config.output_checkpoint_path
\end{lstlisting}

%% ─────────────────────────────────────────────────────────────────────────────
% \subsection{Analysis Utilities}
% \label{sec:appendix-analysis}

% \tamperbench includes utilities for aggregating sweep results and generating visualizations. The epsilon-bounded analysis filters trials by utility degradation and identifies the best attack configurations subject to a constraint---matching the threat model of an attacker who seeks a capably harmful model.

% \newpage

% \begin{lstlisting}[language=bash, label={lst:analysis-usage}]
% # Aggregate sweep results with utility constraints
% python scripts/analysis/analyze_results.py results/sweep epsilon \
%     --epsilons 0.05 0.10 0.20

% # Generate heatmap visualizations
% python scripts/figures/plot_main_heatmap.py results/aggregated_eps05/heatmap.json
% \end{lstlisting}

% \noindent The analysis pipeline produces aggregated metrics, best-configuration YAML files for reproducibility, and JSON exports suitable for visualization scripts.

\end{document}